\documentclass[preprint,prc,a4paper,tightenlines,epsf,nofootinbib,showpacs]{revtex4}
\usepackage{epsfig}

\newcommand{\maM}{\mathcal{M}}
\newcommand{\maA}{\mathcal{A}}
\usepackage{xcolor}
\usepackage{psfrag}

\newcommand{\boldpi}{\mbox{\boldmath $\pi$}}
\newcommand{\boldtau}{\mbox{\boldmath $\tau$}}
\newcommand{\boldT}{\mbox{\boldmath $T$}}

\newcommand{\be}{\begin{eqnarray}}
\newcommand{\ee}{\end{eqnarray}}
\newcommand{\beq}{\begin{equation}}
\newcommand{\eeq}{\end{equation}}

\begin{document}
\preprint{FZJ-IKP-TH-2009-24}
\preprint{HISKP-TH-09/27}
\preprint{ECT*-09-07}

\title{$p$--wave pion production from nucleon-nucleon collisions}
\author{V. Baru$^{1,2}$, E.~Epelbaum$^{1,3}$, J.~Haidenbauer$^{1,4}$,
  C.~Hanhart$^{1,4}$, A.E.~Kudryavtsev$^2$, V.~Lensky$^{5,2}$ 
  and U.-G.~Mei\ss ner$^{1,3,4}$} 

\affiliation{\small $^1$Institut f\"ur Kernphysik
  (Theorie) and J\"ulich Center for 
Hadron Physics, 
Forschungszentrum J\"ulich GmbH, D-52425 J\"ulich, Germany\\
$^2$Institute for Theoretical and Experimental Physics,
 117218, B. Cheremushkinskaya 25, Moscow, Russia\\
$^3$Helmholtz-Institut f\"{u}r Strahlen- und Kernphysik (Theorie), 
 Bethe Center for Theoretical Physics, Universit\"at Bonn, D-53115 Bonn, Germany\\
$^4$Institute for Advanced Simulation, 
Forschungszentrum J\"ulich GmbH, D-52425 J\"ulich, Germany\\
$^5$ European Centre for Theoretical Studies in Nuclear Physics and Related Areas (ECT*),
 Strada delle Tabarelle 286, Villazzano (Trento), I-38050 TN, Italy} 



\begin{abstract}
We investigate $p$-wave pion production in nucleon-nucleon
collisions up to next-to-next-to-leading order
in chiral effective field theory. In particular, we
show that it is possible to describe simultaneously the $p$-wave amplitudes
in the $pn\to pp\pi^-$, $pp\to pn\pi^+$, $pp\to d\pi^+$ channels by 
adjusting a single low-energy constant accompanying the short-range operator
which is available at this order. This study provides a non-trivial test
of the applicability of chiral effective field theory 
to reactions of the type $NN\to NN\pi$.
\end{abstract}


\maketitle

\section{Introduction}

With the advent of chiral perturbation theory (ChPT), the low-energy effective
field theory (EFT) of QCD, high accuracy calculations for
hadronic reactions with a controlled error estimation have become
possible~\cite{chpt1,chpt2}. In that framework,  
$\pi\pi$~\cite{pipi} and $\pi N$~\cite{piN} scattering observables and nuclear
forces~\cite{NN} are calculated based on a perturbative expansion in
$q/\Lambda_{\chi}$ with $q$ referring to  
either a generic momentum of external particles or the pion mass $m_{\pi}$, 
and $\Lambda_{\chi}\sim$ 1 GeV being the chiral symmetry breaking scale. 
An extension of this scheme to pion production in nucleon-nucleon ($NN$) 
collisions turned out to be considerably more difficult. 
A straightforward application of the power counting proposed by Weinberg
\cite{Weinberg:1990rz,Weinberg:1991um} to the reactions $NN\to
NN\pi$~\cite{NNpi,NNpicharged} failed badly (see also Ref.~\cite{BKMnovel} where it was pointed 
out that the naive power counting using the heavy baryon formalism is not applicable above the 
pion production threshold). Indeed, for 
neutral pion production in $pp$ collisions, the corrections due to
the next-to-leading order (NLO) increased the discrepancy with the data and,
moreover, the next-to-next-to-leading order (NNLO) contributions turned out to
be even larger than the NLO terms \cite{NNpiloops}. 
The origin of these difficulties was identified quite early by Cohen et
al.~\cite{bira1}, see also \cite{rocha}, who stressed that the
additional new scale, inherent in reactions of the type
$NN\to NN\pi$, needs to be accounted for in the power counting.
Since the two nucleons in the initial state need to have sufficiently
high kinetic energy to produce the onshell pion in the final state, the
initial center-of-mass momentum needs to be larger than
\begin{equation}
p_{\rm thr} = \sqrt{M_N \, m_\pi}\,,
\quad 
\mbox{with}
\quad 
\frac{p_{\rm thr}}{\Lambda_\chi} \simeq  \ 0.4 \,,
\label{expand}
\end{equation}
where $m_\pi$ and $M_N$ refer to the the pion and nucleon mass, respectively. 
The proper way to include this scale was presented in Ref.~\cite{ch3body}
and implemented in Ref.~\cite{withnorbert}, see Ref.~\cite{report} for a
review article. As a result, pion $p$-wave production is governed by the 
tree-level diagrams up to NNLO in the modified power counting
scheme of Ref.~\cite{ch3body}. On the other hand, for pion $s$-wave production, 
pion loops start to contribute already at NLO.
It was demonstrated in Ref.~\cite{lensky2} that all irreducible 
loop contributions at NLO  cancel altogether,
and the net effect of going to NLO was shown to increase the 
most important operator for charged pion production, first investigated in
Ref.~\cite{koltunundreitan}, by a factor of 4/3. This was sufficient to overcome the
apparent discrepancy with the data in that channel.
But the neutral pion channel is more challenging --- it still calls for a calculation
of subleading loop contributions. First steps in this direction
were taken in Refs.~\cite{subloops}.
We further emphasize that the $\Delta$(1232) isobar should be
taken into account explicitly as a dynamical degree of freedom~\cite{bira1} 
because the Delta-nucleon mass difference, $\Delta M$, is also of the order of $p_{thr}$. 
This general argument was confirmed numerically in phenomenological
calculations~\cite{jouni,ourdelta,ourpols}.

\begin{figure}[t!]
\begin{center}
\includegraphics[scale=1.,clip=]{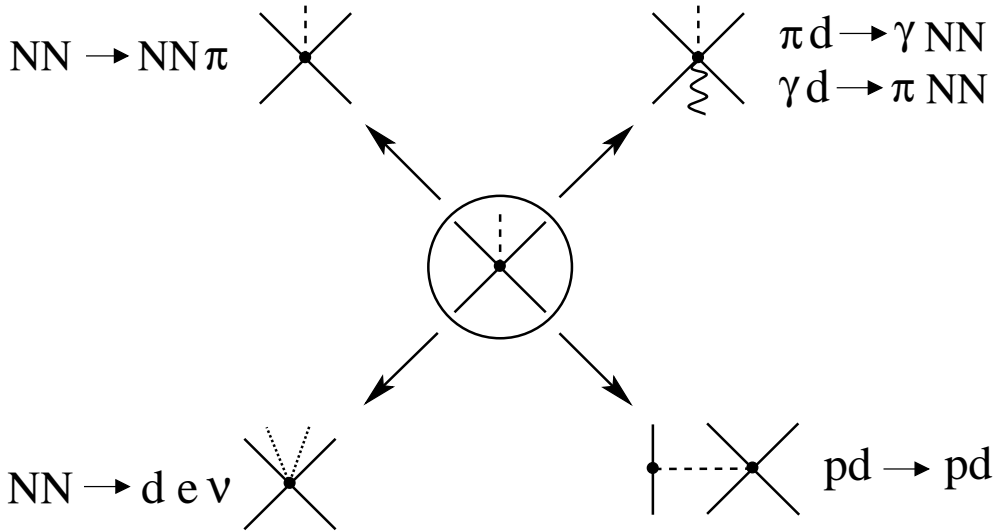}
\end{center}
\caption{\label{4npi}Illustration of the various reactions, where
the leading $(N \bar N)^2 \pi$-contact term contributes.}
\end{figure}

Pion $p$-wave production in $NN$ collisions receives an important
contribution from the leading $(\bar NN)^2\pi$ contact term in the effective
Lagrangian, which also figures importantly in the three-nucleon force
\cite{ch3body,pdchiral}. In addition, the same
operator also contributes to the processes $\gamma d\to\pi
NN$~\cite{gammad,gammad_nn} and $\pi d\to \gamma NN$~\cite{gardestig,gardestig2}
as well as to weak reactions such as, e.g., tritium beta
decay and proton-proton $(pp)$ fusion~\cite{park2,phillips}, as visualized in Fig.~\ref{4npi}. 
Notice that this
operator appears in the above reactions in very different
kinematics, ranging from very low energies for both incoming and
outgoing $NN$ pairs in $pd$ scattering and the weak reactions up to
relatively high initial energies for the $NN$ induced pion production.
In Ref.~\cite{gazit} it was shown that both the $^3$H and $^3$He binding energies
and the triton $\beta$-decay can be described with the same contact term. 
However, an apparent discrepancy between the strength
of the contact term needed in $pp\to pn\pi^+$ and in $pp\to de^+ \nu_e$ 
was reported in Ref.~\cite{nakamura}. If the latter observation were true, it would 
certainly question the applicability of chiral EFT to the reactions $NN\to NN\pi$.

To better understand the
discrepancy reported in Ref.~\cite{nakamura}, in this paper we simultaneously analyze
different pion production channels. In particular, we calculate the
$p$-wave amplitudes for the reactions $pn\to pp\pi^-$, $pp\to
pn\pi^+$, and $pp\to d\pi^+$. Note that even in these channels the contact
term occurs in entirely different dynamical regimes. For
the first channel $p$-wave pion production  goes along with the
slowly moving protons in the ${^1\!}S_0$ final state whereas for the other
two channels the ${^1\!}S_0$ $pp$ state is to be evaluated at the
relatively large initial momentum. Notwithstanding, all three
channels of the reaction $NN\to NN\pi$ seem to give consistent results
for the low-energy constant (LEC) $d$ that represents the strength of the contact term, 
as we will show in the present paper. We discuss which additional data are needed to
further support this conclusion. We argue that 
the origin of the discrepancy reported in Ref.~\cite{nakamura} is not due to  
the different kinematics of $NN\to NN\pi$ and $pp$ fusion, 
but rather in the inconsistency in the partial wave amplitudes used 
in the analysis.
In addition, we also comment on  technical issues related to the work
of Ref.~\cite{nakamura}.

Our manuscript is organized as follows: In Sec.~\ref{general} 
we discuss the general features and the relevant observables for  
$p$-wave pion production. In Sec.~\ref{formalism} the 
power counting is outlined with special emphasis on
the $p$-wave amplitudes. Our
results for the various pion production channels are
presented in Sec.~\ref{results}.
Here, we also discuss the role of the leading $\pi N$
scattering parameters, $c_3$ and $c_4$, for the
$p$-wave pion production amplitudes. We close with a short Summary.

\section{General Remarks}
\label{general}

It is not obvious, a priori, that with just a single contact term,
which contributes to the various reactions shown in Fig.~\ref{4npi},
a consistent description of all these channels can be achieved.
The purpose of the contact term is twofold: it should, on the one hand, 
absorb 
any sensitivities to the employed $NN$ wave functions and in this way 
remove the model dependence in the evaluation of the observables. 
On the other hand, it provides a parameterization of the
short-range physics that contributes to the process being considered.
Thus, the strength of the contact term is necessarily dependent on the
method applied to regularize the integrals (typically a cut-off) and
also on the $NN$ interaction that is used for generating the wave functions.

For the case at hand the contact term connects $NN$ $S$-waves in the initial 
state with 
$NN$ $S$-waves in the final state. 
Since the contact term is a local four-nucleon operator, after 
including the NN distortions its contribution scales as the 
product of the initial and final NN wave functions at the origin.
 Each of these wave functions, in turn, may be represented
by the inverse of the corresponding Jost function~\cite{goldbergerwatson}.
The reason why it is expected to be possible that the same contact term can be used
in all reactions listed above is that the energy dependence of the Jost function is
fixed by the onshell $NN$ data and is therefore independent
of the unknown short range physics. Specifically, the $NN$ distortions can be
represented as an integral 
over the relevant phase shifts by means of the so-called Omn\`es
function~\cite{Omnes} --- see also the discussion in
Ref.~\cite{Ashot}. This is correct up to contributions from the
left-hand cuts and the high energy behaviour of the $NN$ interaction,
both are expected to be of higher order in the expansion. 
As opposed to the energy dependence, the overall scale of the distortions can be shown to be
sensitive to things like the $NN$ interaction and the cut-off
employed~\cite{withkanzo}. Clearly, what needs to be assumed in the
argument given is that there is a proper separation of scales in
the problem. Note that the expansion parameter $p_{\rm  thr}/\Lambda_\chi \sim 0.4$ 
is quite large. In this sense a consistent description of all mentioned reactions with the
same contact term provides a non-trivial test of the applicability of the
chiral expansion to pion production in $NN$ collisions.

One might ask why we take the effort of this study,
since in Ref.~\cite{nakamura} it was already shown
that a consistent description is not possible. 
The answer is twofold: first of all, we found that the
partial wave decomposition of Ref.~\cite{Flammang}, the result of which
was used in Ref.~\cite{nakamura}, is not correct (see discussion in Sec.\ref{pnpppi}) --- this
is why we decided to directly compare to the data in the present work.
Secondly, there is also a conceptual problem in 
the work of Ref.~\cite{nakamura}:
as was outlined above, as long as different phase-equivalent 
$NN$ interactions are used, it should be possible to absorb the model
dependence of the calculation in a single counter term up
to higher-order corrections. 
However, in Ref.~\cite{nakamura} pion production from initial
$NN$ and $N\Delta$ states is not treated on equal footing. 
Rather the contribution from the $\Delta$ isobar excitation
is added on top of and independently of the employed $NN$ 
interactions. 
Thus, it is quite possible that the utilized $NN\to N\Delta$ transition 
potential is too strong. Specifically, it is not constrained by the 
empirical $NN$ phase shifts as it is the case when considering
the $NN$ and $N\Delta$ amplitudes consistently within a coupled-channel 
(Lippmann-Schwinger-like) scattering equation \cite{jouni}. 
In this sense, it should not come as a surprise that it was not possible
to absorb the model dependencies in a single counter term
within the scheme used in Ref.~\cite{nakamura}.
To avoid this problem, in this work we employ the 
coupled-channel $NN$ model of Ref.~\cite{CCF} which involves
the $NN\to N\Delta$ transition potential.

Eventually all reactions shown in Fig.~\ref{4npi} should
be analysed consistently. This would, however, require a calculation to
third order in the chiral expansion of the process $\gamma d\to \pi NN$ and $\pi d\to \gamma NN$
or a rather involved three-nucleon calculation for the tritium beta decay which
goes beyond the scope of this work. Instead, as a next step in this ambitious
program, we analyse here in detail various pion production channels. Notice
that although these reactions appear to have  similar kinematics, the relevant
transition for the reaction $pn\to pp\pi^-$ involves very low momenta in
the $NN$ $^{1\!}S_0$ state and considerably higher momenta 
($\sim p_{\rm thr}$) in the
$^{3\!}S_1$ channel while the situation is just opposite for the reactions
$pp\to (d/pn)\pi^+$. Thus, a simultaneous description of these reaction with a
single short-range operator indeed provides a highly nontrivial consistency
test of our approach. 
Notice further that the $(N \bar N)^2 \pi$ short-range operator we are interested in
here does not contribute to the $pp\to pp\pi^0$ transition which is,
therefore, not considered in the present work. Thus, the only reactions of
interest for this study are $pp\to (pn/d) \pi^+$ and
$pn\to pp\pi^-$. Here the relevant transitions
are ${^{1\!}}S_0\to {^{3\!}}S_1p$ for the former
and  $({^{3\!}}S_1-{^{3\!}}D_1)\to {^{1\!}}S_0p$ for the
latter, where the small letter labels the pion angular
momentum. Since the main focus of this work is on the role
of the contact term, we will concentrate on observables
where the final $NN$ system is in an $S$-wave --- which
largely simplifies the numerical work.
However, as outlined below, the contribution 
of $NN$ $P$-waves to observables might be relevant for the reaction
$pp\to pn\pi^+$. This potential problem renders this channel
not very convenient for the extraction of the counter term, as will be
discussed in Sec.~\ref{results}.
 
To be specific, we calculate in this work the
differential cross sections and analyzing powers for the reactions 
$pp\to d\pi^+$,  $pp\to pn\pi^+$, and $pn\to (pp)_{^{1\!}S_0}\pi^-$.
Here the symbol $(pp)_{^{1\!}S_0}$ indicates that 
in the corresponding measurement the final $pp$ relative
momentum was restricted kinematically to be less than 38 MeV/c  ($M_{pp}-2M_N \le 1.5$~MeV) 
which leads to a projection on the ${^{1\!}}S_0$ $pp$ final state.
For all the mentioned observables experimental data are available
or will be available soon in the energy range of relevance here.
Besides the anisotropy of the pion angular distributions, all observables are
sensitive to both $s$- and $p$-wave pion
production. Although there exists an NLO calculation
for $s$-wave pion production in $pp\to d\pi^+$
using ChPT, its theoretical uncertainty is still sizable~\cite{lensky2}.
For $s$-wave pion production accompanied by a
transition of an isospin-one $NN$ pair to an isospin-one $NN$ pair (e.g. in $pp\to pp\pi^0$),
no sufficiently accurate ChPT calculation
is available at present. Since we focus here on the $p$-wave amplitudes,
we extract the $s$-wave amplitudes directly
from the data in order to minimize the uncertainties of our calculation.
The phase of these amplitudes is then imposed using the Watson
theorem~\cite{goldbergerwatson}, see the discussion in Sec.~\ref{results}.

It is well known that $p$-wave pion production
in  $pp\to d \pi^+$ and in $pp\to pn \pi^+$ is strongly dominated
by the transition ${^{1\!}}D_2\to {^{3\!}}S_1p$ due to a strong
coupling of the initial $NN$ state to the ${^{5\!}}S_2$ $N\Delta$ state~\cite{ericsonweise}.
Therefore, the amplitude we are interested in has only a minor impact on the
observables. In other words, the uncertainty for the extraction of the counter
term from these reactions will be significant. The situation is much more
promising for the reaction $pn\to pp\pi^-$:
here the amplitude of interest is the leading $p$-wave.
In addition, the strength of the $s$-wave amplitude can be
taken from the reaction $pp\to pp\pi^0$ using isospin symmetry and correcting
for the final state interaction (FSI) as discussed in Sec.~\ref{pppnpi}. 
Unfortunately, no data are presently available for $pn\to pp\pi^-$ at sufficiently low
energies. Nevertheless, as will be shown below, already the higher-energy
data provide some insights. In addition,
data at lower excess energies will be available soon~\cite{ANKE}.

The goal of the present investigation is to explore whether it is possible to
obtain a simultaneous description of all $NN\to NN\pi$ channels. 
A more quantitative study including a statistical
analysis of the data and an estimation of the theoretical uncertainty is postponed
until accurate experimental data will become available for $pn\to pp\pi^-$ at
low energies. 

\section{Formalism}
\label{formalism}

Our calculations are based on the effective chiral Lagrangian with explicit
$\Delta$ degrees of freedom. The leading $\pi N$ and $\pi N \Delta$
interaction terms read 
\cite{OvK,ulfbible}
\begin{equation}
 {\cal L}^{(0)}  = 
 \nonumber
   N^{\dagger}\left[\frac{1}{4 f_{\pi}^{2}} \boldtau \cdot
         (\dot{\boldpi}\times{\boldpi})
         +\frac{g_{A}}{2 f_{\pi}}
         \boldtau\cdot\vec{\sigma}\cdot\vec{\nabla}\boldpi
\right]N 
+\frac{h_{A}}{2 f_{\pi}}\left[N^{\dagger}(\boldT\cdot
          \vec{S}\cdot\vec{\nabla}\boldpi)\Psi_\Delta +h.c.\right] +\cdots \ ,
\label{la0}
\end{equation} 
while the first corrections  have the form 
\begin{eqnarray}
 {\cal L}^{(1)}&=&
    \frac{1}{8M_N f_{\pi}^{2}}
    (iN^{\dagger}\boldtau\cdot
        (\boldpi\times\vec{\nabla}\boldpi)\cdot\vec{\nabla}N + h.c.)
                                       -\frac{1}{f_{\pi}^{2}}N^{\dagger}\bigg[ c_3 (\vec{\nabla}\boldpi)^{2}
       +
        \frac{1}{2} \left(c_4 + \frac{1}{4M_{N}}\right) \nonumber \\
&\times&        \varepsilon_{ijk} \varepsilon_{abc} \sigma_{k} \tau_{c}
        \partial_{i}\pi_{a}\partial_{j}\pi_{b}\bigg]N 
 -\frac{d}{f_{\pi}}
        N^{\dagger}(\boldtau\cdot\vec{\sigma}\cdot\vec{\nabla}\boldpi)N\,
        N^{\dagger}N +\cdots  \, \ .            \label{la1}
\end{eqnarray}
\noindent
The ellipses stand for further terms which are not relevant for the present
study. In the equations above $f_\pi$ denotes the pion decay constant in the chiral limit,
 $g_A$ is the axial-vector coupling of the nucleon, $h_A$ is the $\Delta N \pi$
coupling, $N$ and $\Psi_\Delta$ correspond to 
the nucleon and Delta fields, respectively, and
$\vec S$ and $\boldT$ are the transition spin and isospin
matrices, normalized according to:
\begin{equation}
S_iS_j^\dagger 
=
\frac{1}{3}(2\delta_{ij}-i\epsilon_{ijk}\sigma_k) \ ,\quad \quad
T_iT_j^\dagger =
 \frac{1}{3}(2\delta_{ij}-i\epsilon_{ijk}\tau_k) \ .
\label{deltanorm}
\end{equation}
 We also emphasize that the effective Lagrangian of Ref.~\cite{OvK}
contains another $(\bar NN)^2\pi$ contact operator which can be shown to be redundant
as a consequence  of the Pauli principle
\cite{ch3body,pdchiral,park2}.

\begin{figure}[tb]
\begin{center}
\includegraphics[scale=0.5,clip=]{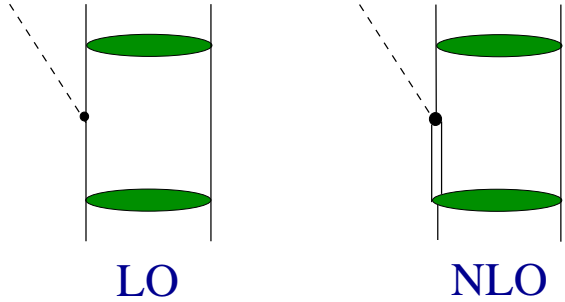}
\end{center}
\caption{\label{fig:NN2NNpi_direct} "(Color online)" Leading 
and next-to-leading 
diagrams for the $p$-wave amplitudes of $NN\to NN\pi$. Single (double) solid lines denote
nucleons (Deltas), dashed lines denote pions, green ellipses correspond to the $NN$ wave functions in 
the initial and final states. }
\end{figure}

\begin{figure}[tb]
\begin{center}
\includegraphics[scale=0.7,clip=]{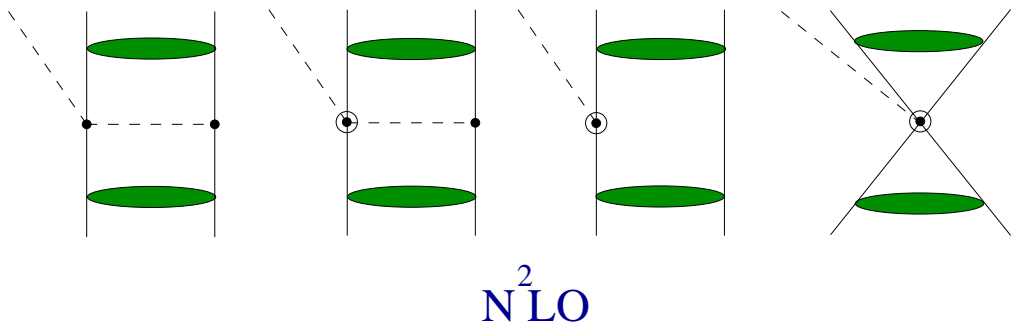}
\end{center}
\caption{\label{fig:NN2NNpi_NNLO}  "(Color online)"
Diagrams that contribute at NNLO to
the $p$-wave amplitudes of $NN\to NN\pi$.
Subleading vertices are marked as $\odot$. 
}
\end{figure}

We are now in the position to 
discuss the relevant scales and counting rules for $p$-wave pion
production.  We assign the outgoing two-nucleon relative momentum $
p'$ and the outgoing pion momentum $k_{\pi}$ to be of order of $ m_{\pi}$
and introduce the expansion parameter \be
\chi\simeq\frac{k_{\pi}}{p}\simeq\frac{p'}{p}\simeq\frac{m_{\pi}}{p}\simeq\frac{p}{M_N}\simeq\frac{\Delta M}{M_N}
\ee where $p\simeq p_{\rm thr}$ is the initial two-nucleon relative
momentum. The counting rules for the time-dependent vertices,
such as e.g.~the Weinberg-Tomosawa (WT) vertex in $
{\cal L}^{(0)}$, are discussed in detail in
Refs.~\cite{lensky2,NNpiMENU}. At leading order one finds that the WT
vertex is $\propto 2\omega_\pi$  with
$\omega_\pi$ being the energy of the outgoing (onshell) pion. 
The diagrams contributing to the production operator at LO and at
NLO are shown in  Fig.~\ref{fig:NN2NNpi_direct} whereas the corresponding graphs at
NNLO are depicted in Fig.~\ref{fig:NN2NNpi_NNLO}. 
At NLO there are
only diagrams in which the pion is produced through the excitation of
the $\Delta$ resonance. The relative suppression of these diagrams as
compared to the ones involving the nucleon is accounted for by the $\Delta$
propagator which is suppressed by $1/p$ as compared to $1/m_{\pi}$ in
the nucleon case.
To see that the diagrams in Fig.~\ref{fig:NN2NNpi_NNLO}
indeed contribute at NNLO for $p$-wave pion production consider, as an
example, the first graph in this figure. Its contribution can be estimated
using dimensional analysis as follows:
\be
\frac{\omega_\pi}{f_{\pi}^2}\frac{1}{p^2}\frac{k_{\pi}}{f_{\pi}}\simeq
\frac{1}{f_{\pi}^3} \frac{k_{\pi}}{m_{\pi}} \frac{m_{\pi}}{M_N}.
\ee
Here we used that the outgoing pion
momentum $k_{\pi}$ enters the $\pi NN$ vertex to allow for the  $p$-wave amplitude.
To understand the suppression factor this operator should be
compared with the LO contribution ${k_{\pi}}/({f_{\pi}^3}{m_{\pi}}) $.
Thus, one gets an order $\chi^2$ suppression for the first diagram of
Fig.~\ref{fig:NN2NNpi_NNLO}. Similarly, using the
$\pi\pi NN$ vertex from $ {\cal L}^{(1)}$  
in combination with the $p/f_{\pi}$-scaling for the $\pi NN$ vertex one arrives again at
a $\chi^2$ suppression for the second diagram in  Fig.~\ref{fig:NN2NNpi_NNLO}.
Further details can be found in Appendix~\ref{amplitudes} which
contains explicit expressions for the diagrams shown in
Figs.~\ref{fig:NN2NNpi_direct}  and~\ref{fig:NN2NNpi_NNLO}.
Once the amplitudes are evaluated they need to be convoluted with
proper $NN$ wave functions. Ideally, one would use wave
functions derived from the same formalism, namely ChPT. However, up to
now these are only available for energies below the pion production 
threshold~\cite{NN}. We therefore use
the so-called hybrid approach, first introduced by Weinberg \cite{Weinberg:1991um}, based on
the transition operators derived within the effective field theory and
convoluted with realistic wave functions~\cite{CCF}. This procedure
should also provide reasonable results, however, a reliable
uncertainty estimate is possible only at the level of the transition
operator.

\section{Results and Discussion}
\label{results}

\subsection{Parameters of the calculation}

To the order we are working, the following low-energy constants (LECs) appear 
in the calculation: $f_\pi$, $g_A$, $h_A$, $c_3$, $c_4$ and $d$.  Only the
last LEC cannot be taken from other sources, for its value 
strongly depends on the $NN$ wave functions employed. 
We adopt the following values of the parameters: $f_{\pi}=92.4$~MeV,
$g_A=1.32$, $h_A \simeq 2.1 g_A = 2.77$, $c_3=-0.79$~GeV$^{-1}$ and
$c_4=1.33$~GeV$^{-1}$.  The values of the LECs $c_3$
and $c_4$ are taken from Ref.~\cite{Evgeny}. From the fit to $\pi N$ threshold
parameters, two solutions for the $c_i$ are given in Ref.~\cite{Evgeny}
corresponding to the different choices  of
$h_A$ ($h_A \simeq 2.1 g_A$ and $h_A \simeq 2.1$).
The sensitivity of the results to the different values of $c_3$ and $c_4$ will be also
discussed. As already mentioned in the Introduction, the power
counting scheme calls for a dynamical treatment of the $\Delta$
isobar as a result of the  comparable numerical value of the 
Delta-nucleon mass difference and $p_{\rm thr}$. The implications of
integrating out the $\Delta$ degrees of freedom for the processes at hand are
discussed  in Appendix~\ref{saturation}.

The deuteron wave function and the $NN$ scattering amplitudes used
in the calculation are generated from the CCF $NN$ potential~\cite{CCF}.
As described above, we do not calculate the $s$-wave
pion amplitudes in this work but rather take both their strength and the phases
directly from experiment. To be specific, for the reaction $pn\to
pp\pi^-$ we aim at the description of the double differential cross
sections and the analyzing power measured at TRIUMF~\cite{hahn,duncan} and PSI~\cite{Daum}.
Following the Watson theorem to parameterize the relevant ${^{3\!}}P_0\to
{^{1\!}}S_0s$ amplitude, we use the ansatz $\tilde C e^{i\delta_{{}^{3\!}P_0}}
\Psi^{(+)}_{p'}(r=0)$, where the inverse Jost function in the ${^{1\!}}S_0$ partial
wave, $\Psi^{(+)}_{p'}(r=0)$, and the initial phase shift
$\delta_{^{3\!}P_0}$ are calculated from the $NN$ model used, and
the parameter $\tilde C$ is fitted to reproduce the corresponding amplitude
extracted from the TRIUMF data using a partial wave analysis~\cite{duncan,duncan2}.
It is interesting to note that the ${^{3\!}}P_0\to
{^{1\!}}S_0s$ amplitude from the TRIUMF analysis at\footnote{Traditionally,
the energy in the pion production reactions is given in terms of $\eta$,
the (maximum) pion momentum allowed in units of the pion mass.}
$\eta$=0.66 ($T_{lab}=353$~MeV) is about 25\% larger than that extracted 
from the $pp\to pp\pi^0$ measurement at CELSIUS \cite{bilger}.  A similar
inconsistency is discussed in Ref.~\cite{cosytof} where it is argued
that the total cross sections at low energies for $pp\to pp\pi^0$
recently measured at COSY are about 50\% larger than those at CELSIUS
and IUCF as a result of the
missing acceptance at small angles for both CELSIUS and IUCF.

For the reaction $pp\to d\pi^+$ the $s$-wave amplitude occurs in the
${^{3\!}}P_1\to {^{3\!}}S_1s$ partial wave and can be related to the total
cross section at threshold. The most precise way of getting this
quantity is to extract it from the width of pionic deuterium atom,
measured at PSI with high accuracy \cite{Hauser,Strauch}. This
procedure gives the following value of $\alpha$, the total cross section
divided by $\eta$:
$\alpha =252^{+5}_{-11}\ \mu$b~\cite{pidexp}.
Thus, we adjust the magnitude of the ${^{3\!}}P_1\to {^{3\!}}S_1s$ amplitude
  to be in agreement with this observable.

As mentioned above the value of $d$ depends on the $NN$ interaction employed
and on the method used to regularize the overlap
integrals. Indeed, in Refs.~\cite{park2,ch3body}
a strong sensitivity of the LEC $d$ to the
regulator is reported. It therefore does not make
much sense to compare values for $d$ as found in
different calculations. What makes sense, however, is
to compare results on the level of observables and
this is what we will do below. We will adjust the value of $d$  in such a way to get the 
best simultaneous qualitative description of all channels of $NN\pi$.

\subsection{Reaction {\boldmath$pp\to d\pi^+$}}

\begin{figure}[t!]
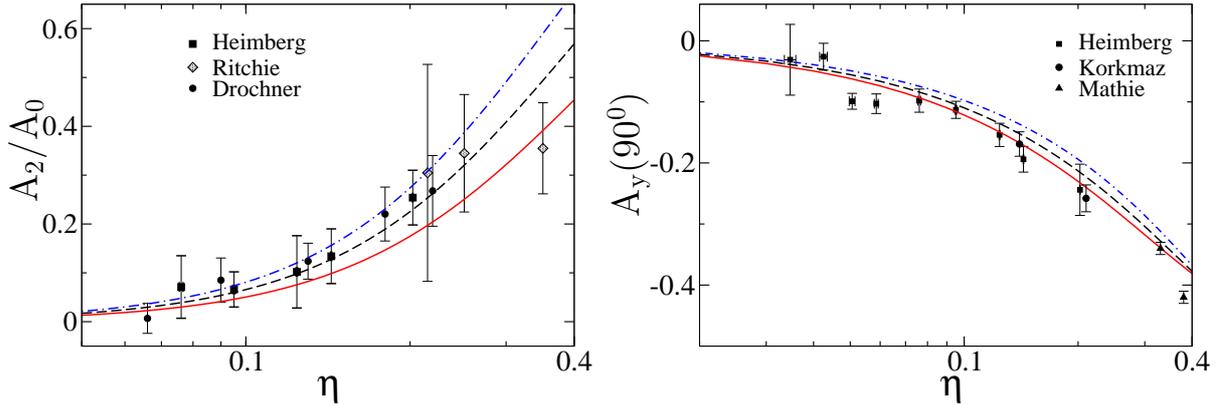

\begin{center}
\psfrag{A2/A0}{\large $\rm A_2/A_0$}
\includegraphics[scale=0.3]{Fig4a.eps}
\hspace{0.1cm}
\psfrag{Ay (90)}{\large $\rm A_y(90^0)$}
\includegraphics[scale=0.3,clip=]{Fig4b.eps}
\end{center}
\caption{\label{fig:Ay90A2oA0} "(Color online)"
Results for $A_2/A_0$ (see Eq. \ref{diff}) (left panel) and  the analyzing power at 90 degrees
(right panel) for the reaction $pp\to d\pi^+$ for different
values for the strength of $d$ (in units  $1/(f^2_{\pi}M_N)$). Shown are $d=3$ (red solid line), 
$d=0$ (black dashed line), and $d=-3$ (blue dot-dashed line). 
The data are from Refs.~\cite{Ritchie,Heimberg,Drochner,Korkmaz,Mathie}.
The strength and phase of the $s$-wave amplitude is fixed from
data.}
\end{figure}

\begin{figure}[t!]
\psfrag{Ay}{\large $\rm A_y$}
\parbox{7cm}{\hspace{-1cm}\includegraphics[scale=0.52,clip=]{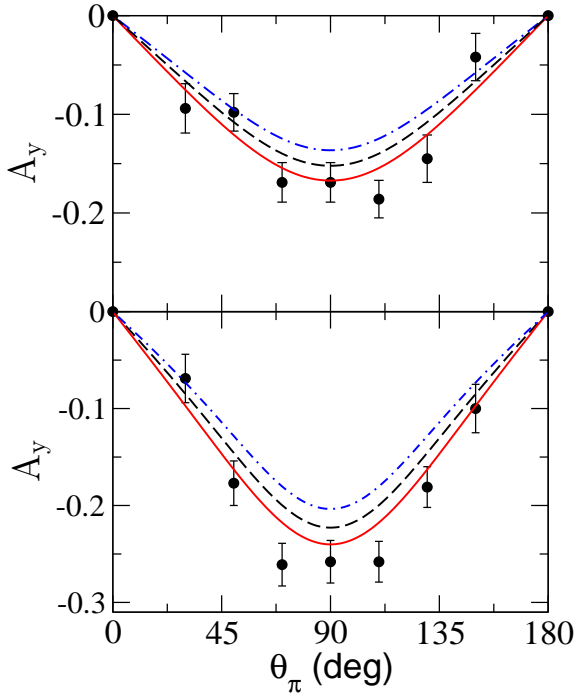}}
\parbox{5cm}{
\caption{\label{fig:Ayotheta}  "(Color online)"
Results for the analyzing power at $\eta$=0.14 (upper panel) and
$\eta$=0.21 (lower panel) as functions of the angle $\theta_{\pi}$
for the reaction $pp\to d\pi^+$ for different
values of $d$. Shown are $d=3$ (red solid line), $d=0$ (black dashed line),
and $d=-3$ (blue dot-dashed line). The data are from Ref.~\cite{Korkmaz}.
The strength and phase of the $s$-wave amplitude is fixed from
data.}}
\end{figure}

We begin with a discussion of the results for
the reaction $pp\to d\pi^+$. In Fig.~\ref{fig:Ay90A2oA0}, 
we compare our calculation for various
values of $d$ with the experimentally available 
angular asymmetry parameter $A_2/A_0$.
The coefficients $A_i$ are related to the
unpolarized differential cross section via
\be
\frac{d\sigma}{d\Omega} = A_0 + A_2 P_2(\cos \theta_\pi) \ ,
\label{diff}
\ee
with $P_2(x)$ being the second Legendre polynomial and $\theta_\pi$ the pion 
angle in the c.m.~frame. We also show results for
the analyzing power at 90 degrees. In both cases the observables are plotted 
as functions of the parameter $\eta$. 
Here and in what follows, the value of the LEC $d$ is always given in units 
$1/(f^2_{\pi}M_N)$. Notice that at low
energies, it is sufficient to just show the analyzing
power at 90 degrees since its angular dependence is
proportional to  $\sin \theta_{\pi}$.
To illustrate this 
we also present in Fig.~\ref{fig:Ayotheta} the analyzing power
as a function of the scattering angle for two different energies 
$\eta$=0.14 and $\eta$=0.21. At $\eta\simeq 0.5$ the angular dependence of
the analyzing power starts to deviate significantly
from $\sin \theta_{\pi}$ due to the onset of $d$-waves.
Clearly, at these (and higher) energies we cannot expect
our calculation to agree with the data anymore.
As can be seen from  the figures, the data
at small $\eta$, especially the analyzing power,  prefers a positive value for $d$ --- our
fit resulted in $d=3$ for the best value.
To demonstrate the effect of the LEC $d$ on the observables, 
in Fig.~\ref{fig:Ayotheta} and in subsequent figures we also give the results
with $d=0$ and with the negative LEC $d=$-3.

\subsection{Reaction {\boldmath$pn\to pp\pi^-$}}
\label{pnpppi}

\begin{figure}[t!]
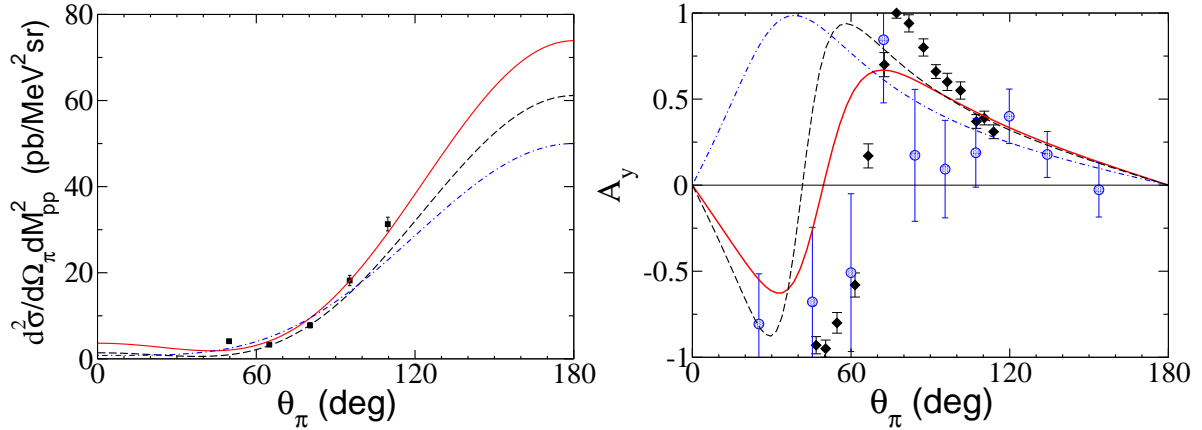

\hspace*{-0.9cm}
\includegraphics[width=0.47\textwidth,clip=]{Fig6a.eps}\hspace*{-0.12cm}
\includegraphics[width=0.475\textwidth,clip=]{Fig6b.eps}
\caption{ "(Color online)" 
Results for $d^2\sigma/d\Omega_{\pi}dM^2_{pp}$ (left panel) and $A_y$
(right panel) for $pn\to pp(^1S_0)\pi^-$. 
Shown are the results for $d=3$ (red solid line), $d=0$ (black dashed line) and $d=-3$
(blue dot-dashed line). 
The data is from  TRIUMF~\cite{hahn,duncan} (black squares) and from
PSI~\cite{Daum} (blue circles) .} 
\label{res_pppimin}     
\end{figure}

We now turn to the reaction $pn\to
pp\pi^-$. As it was already explained in Sec.~\ref{general}, for this reaction channel 
the relevant pion $p$-wave  occurs in conjunction with the two-nucleon pair in the 
${^{1\!}}S_0$ state. It is known experimentally that in the isospin-one channel 
the final $P$-wave diproton contributions ($Pp$ and $Ps$) start growing with
the energy rather rapidly so that already for excess energies around 30 MeV 
they provide about 50\% of the total cross section~\cite{bilger}. 
Therefore, in order to be sensitive to our particular amplitude
one needs to isolate  experimentally the $S$-wave diproton state by putting
kinematical cuts on the two-nucleon relative momentum. This is exactly 
what was done in the experimental study of $pn\to pp\pi^-$ at
TRIUMF~\cite{hahn,duncan}. In particular, they measured the
differential cross section $d^2\sigma/(d\Omega dm_{pp}^2)$ and analyzing power
$A_y$ for $T_{lab}=353$~MeV ($\eta$=0.66), where the final diproton relative
momentum $p'$ was restricted to be not larger than $38$~MeV/c ($M_{pp}-2M_N\simeq 1.5$~MeV). 
 A similar measurement for the analyzing power was also performed at
PSI~\cite{Daum} for $T_{lab}=345$~MeV and $pp$ 
invariant masses $M_{pp}-2M_N < 6$~MeV. It is interesting to note that the
positions of the peaks in $A_y$ seem to be somewhat different in these experiments
(see Fig.~\ref{res_pppimin}), although the data of Ref.~\cite{Daum} have much
larger uncertainties than those of Ref.~\cite{hahn}. 
Unfortunately, presently data for $pn\to pp\pi^-$ are only available
at such high energies 
where our corresponding results in the $pp\to d\pi^+$ 
channel already start to deviate considerably from the experiment.
Therefore, for the reaction $pn\to pp\pi^-$ we expect likewise
only a qualitative description.
Nevertheless, a comparison with the experimental data
in this channel is quite instructive too and shows also a preference 
for a positive value of $d$ as visualized in Fig.~\ref{res_pppimin}. 
Fortunately, there will soon be a measurement
for the same observables at lower energies
at COSY~\cite{ANKE}. Once this data will be available
we should be able to draw more quantitative conclusions
on the value of the parameter $d$ needed for the
reaction $pn\to pp\pi^-$.

\subsection{Reaction {\boldmath$pp\to pn\pi^+$}}
\label{pppnpi}

The reaction $pp\to pn\pi^+$ is the most difficult and the
least convenient one for the extraction of the contact term.  Besides the
fact that here, as in $pp\to d\pi^+$, pion $p$-wave production is
mainly driven by the ${^{1\!}}D_2$ initial state, in addition 
 $NN$  $P$-waves contribute for isospin-one 
as well as for isospin-zero $NN$~final states. 
At the energies considered in the
experimental investigation, $\eta=$0.22, 0.42, and 0.5, the $Pp$
amplitudes may contribute significantly~\cite{bilger,complete,pwpi0}.
They should be particularly important in view of the smallness of the
${^{1\!}}S_0$ amplitude --- even small contributions to $A_2$, see
Eqs.~(\ref{diff}) and~(\ref{A_2}), can affect the partial wave
analysis considerably.  In the partial wave analysis performed in
Ref.~\cite{Flammang}, these $Pp$ contributions were not taken into account at
all. Also there are contributions to $A_0$ from the isospin-one $NN$~final
states that potentially increase the uncertainty of the analysis,
especially in view of the differences in the experimental results in
$pp\to pp\pi^0$ as already discussed above. These arguments alone 
cast serious concerns on the partial wave analysis performed
in Ref.~\cite{Flammang}. But there is an even more direct evidence 
of problems with the extraction of the partial wave amplitudes of
Ref.~\cite{Flammang} which we now discuss in detail.
The observables measured for the reaction ${\vec p} p\to pn\pi^+$ in
Ref.~\cite{Flammang} include the coefficients $A_0$ and $A_2$
in the differential cross section, see Eq.~(\ref{diff}) and the analyzing
power $Ay(90^\circ)$. Neglecting the $Pp$ contributions these
observables can be expressed in terms of the three partial wave
amplitudes with the isospin-zero $pn$-state  $a_0$~(${^{1\!}}S_0\to {^{3\!}}S_1p$) 
 (the single amplitude, where the $(N \bar N)^2 \pi$ contact term contributes), 
$a_1$~(${^{3\!}}P_1\to {^{3\!}}S_1s$), $a_2$~(${^{1\!}}D_2\to {^{3\!}}S_1p$)
and the contribution of the isospin-one channel denoted as $A_0^{I=1}$ via
\be\label{A_0}
A_0&=& \frac{|a_0|^2+|a_1|^2+|a_2|^2}{4}+A_0^{I=1},\\
\label{A_2}
A_2&=& \frac{|a_2|^2}{4} -\frac{1}{\sqrt{2}}{\rm Re}[a_0a_2^*],\\
\label{A_N0}
A_{y}(90^\circ)\left(A_0-\frac{A_2}{2}\right)&=& \frac{1}{4}
(\sqrt{2}{\rm Im}[a_1a_0^*] +{\rm Im}[a_1a_2^*]),  
\ee 
where $(A_0-{A_2}/{2})$ is just ${d\sigma}/{d\Omega}(90^\circ)$ from Eq.~(\ref{diff}).
Using the
system of Eqs.~(\ref{A_0})-(\ref{A_N0}) one can determine the
amplitudes $a_0,\ a_1$ and $a_2$ provided one knows the isospin-one
piece $A_0^{I=1}$. The latter was extracted in Ref.~\cite{Flammang}
from the measurement of the total cross section in the reaction $pp\to
pp\pi^0$ reported in Ref.~\cite{Meyer}. However, the FSI in the
$pp\to pp\pi^0$ reaction is very different to that in the $pp\to
pn\pi^+$ channel. To estimate the difference note that in the energy
region studied, which is less than $20$~MeV, the dominant partial wave
is the one where the final two-nucleon state is in the $S$-wave. In
this case the correction factor would be proportional to the ratio of
the inverse Jost functions squared integrated over the phase space
\be
R=\frac{\int d^3 p' k_{\pi} |F_\mathrm{pn}(p')|^2}{\int d^3 p' k_{\pi}
  |F_\mathrm{pp}^{\rm CC}(p')|^2} \ , 
\ee
where $F_\mathrm{pn}(p')$ and $F_\mathrm{pp}^{\rm CC}(p')$ are the
inverse Jost functions for the $pn$ and $pp$ ${^{1\!}}S_0$ states, respectively.  As
discussed above, although the Jost function itself depends on the
$NN$ model used, its energy dependence does not. We may, therefore,
evaluate $R$ using any sensible model for the $NN$ interaction.  For a 
separable $NN$ potential there exists an analytic
expression for the Jost function in the $pp$ system in the presence of
the Coulomb interaction~\cite{vanher,KDT}. Using it one finds that the
ratio $R$ is about 1.5 for $\eta=0.22$ and about 1.2 for $\eta=0.42$.
Similar results are obtained using the CCF $NN$ interaction \cite{CCF}.
Thus, compared to the original analysis performed in
Ref.~\cite{Flammang}, the isospin-one contribution at $\eta=0.22$ should 
be enhanced by more than a factor of two if, in addition, one utilizes the
new, larger experimental data from COSY for the total cross section
for $pp\to pp\pi^0$~\cite{cosytof}.
This change, of course, will significantly affect the
results of the partial wave analysis.
Given the above difficulties with the partial wave analysis of
Ref.~\cite{Flammang}, we decided to compare our
results directly to the experimentally measured quantities.
Aiming presently at a qualitative description of the data,
we will not include the $Pp$-states in this work. 

\begin{figure}[tb]
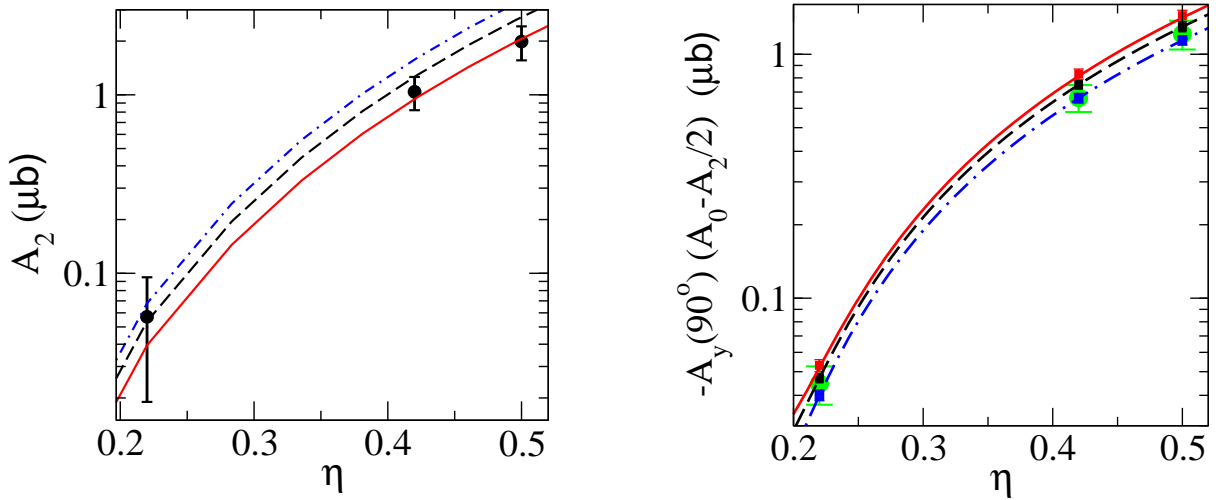

\parbox{0.47\textwidth}{\includegraphics[scale=0.58,clip=]{Fig7a.eps}}
\hfill
\parbox{0.47\textwidth}{\includegraphics[scale=0.55,clip=]{Fig7b.eps}}
\caption{\label{A2}  "(Color online)"
Results for the magnitude of $A_2$ (left panel)
and  $A_y(90^\circ)(A_0-A_2/2)$  (right panel) for the reaction $pp\to pn\pi^+$
for different values of the contact term. 
The notation of curves is the same as in Fig.~\ref{fig:Ay90A2oA0}. 
The data are from Ref.~\cite{Flammang}.} 
\end{figure}

The results of our calculation for $A_2$ are shown in Fig.~\ref{A2}. Again,
positive values of the contact term with $d \sim 3$ seem to be preferred.
We emphasize, however, that these results should be treated with care since
the calculations at higher energies can be affected by $P$-wave
contributions whereas the lowest point is not very sensitive to the
value of $d$ due to the large experimental uncertainty.  We can
also check whether our results are consistent with the measurement of
the analyzing power that is related to our amplitudes via
Eq.~(\ref{A_N0}). To allow for this comparison, however, we need to
know the pion $s$-wave amplitude $a_1$. At present, this quantity is known
theoretically only up-to-and-including terms
at NLO. Therefore, to minimize the uncertainty of the current study, we
extract this amplitude directly from data on the total cross section
in $pp\to pn\pi^+$ through Eq.~(\ref{A_0}). We employ the amplitude
$A_0^{I=1}$ consistent with the
data at COSY and correct for the FSI factor as described above and
take the amplitudes $a_0$ and $a_2$ from our NNLO calculation.
In the right panel of Fig.~\ref{A2} we compare our results for
$A_{y}(90^\circ)\left(A_0-{A_2}/{2}\right)$ with the corresponding
data. Since we use the experimental total cross sections
to extract $a_1$, our results, given by red %
($d=3$), black ($d=0$) and blue ($d=-3$) squares
in the right panel of Fig.~\ref{A2}, can be presented at specific energies only. 
The squares include the experimental uncertainty in the total cross section $A_0$ used to extract $a_1$.
To guide the eye, we also show the results of interpolations
between the three energies. It is seen that 
the magnitude  $A_{y}(90^\circ)\left(A_0-{A_2}/{2}\right)$ is much less sensitive 
to the value of the LEC $d$ than, e.g., $A_2$.  
Notice further that the experimental points do not include a 12\% uncertainty
due to systematic errors in $A_0$ and $A_2$.

\begin{figure}[t!]
  \psfrag{a0}{\hspace*{-1.1cm}\Large $\rm\ a_0\  (\, {{\mu b}^{1/2}}\, )$}
\parbox{7cm}{\hspace{-1cm}\includegraphics[scale=0.52,clip=]{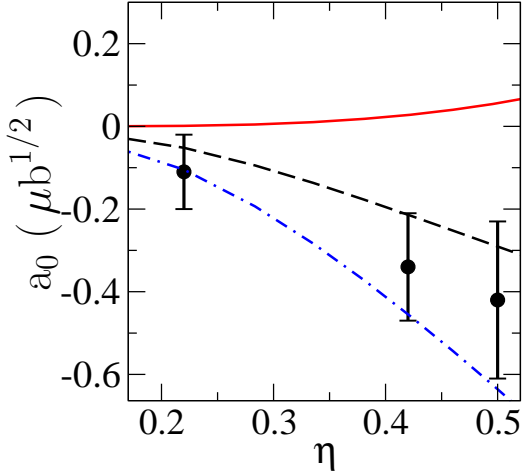}}
\parbox{5cm}{
\caption{\label{a0Fig}  "(Color online)"
Comparison of $a_0$ as it results from our analysis in comparison
to the partial wave amplitude extracted in Ref.~\cite{Flammang}. 
The notation of curves is the same as in Fig.~\ref{fig:Ay90A2oA0}. 
Since our
calculations well describe all observables of Ref.~\cite{Flammang}, this figure
nicely illustrates the problem of the partial wave decomposition of this reference.}}
\end{figure}

Since we do not know the contribution of the $NN$ $P$-waves
to the $pp\to pn\pi^+$ observables at present, and an improved partial wave analysis
would require a careful study of various uncertainties, we do not
try to extract $a_0$ from the data. However, in order to illustrate the
potential
effect of the changes discussed above (up to $NN$ $P$-waves) on $a_0$, in
Fig.~\ref{a0Fig} we show the results of our calculation for $a_0$ in
comparison to the old extraction of Ref.~\cite{Flammang}. 
Evidently, although all data presented in Ref.~\cite{Flammang} 
are in a good agreement with our calculation (as demonstrated in
Fig.~\ref{A2}), the partial wave amplitude is not at all described --- 
see solid curve for our results with $d$=3 in Fig.~\ref{a0Fig},
which illustrates clearly that the partial wave solution
given in Ref.~\cite{Flammang} should be abandoned. 
It is interesting to note that in Ref.~\cite{nakamura} it was stressed
that a positive value for $a_0$ is necessary in order to achieve a result for 
pion production that is consistent with the ones for the weak rates. 
This is in accord with our findings based solely on the data for $NN\to NN\pi$. 
Here we do not aim at a more quantitative comparison with Ref.~\cite{nakamura},
because of the technicalities discussed in the beginning of Sec.~\ref{general}.

Finally, we would like to discuss the sensitivity of
our results to the parameters $c_i$.  As shown in
Appendix~\ref{amplitudes}, for the ${^{1\!}}S_0\to
{^{3\!}}S_1$ or ${^{3\!}}S_1\to {^{1\!}}S_0$ $NN$ transitions the parameters $c_i$ occur in the
combination $C_i^{^{3\!}S_1}=c_3/2+c_4+1/(4M_N)$.  This combination
appears to be largely constrained by the $\pi N$ data since the
different sets of $c_i$ from the recent analysis~\cite{Evgeny} give
basically the same value for $C_i^{^{3\!}S_1}$.
In addition, to
the order we are working at, this combination is fully absorbed in the
counter term since the corresponding potential for $NN\to NN\pi$, see
the second diagram  in Fig.~\ref{fig:NN2NNpi_NNLO}, is just a constant up to
higher order terms 
\be V^{c_i}_{^{1\!}S_0,^{3\!}S_1} \sim  C_i^{^{3\!}S_1}\frac{(\vec p- \vec
  p\,')^2}{(\vec p- \vec p\,')^2+m_{\pi}^2} = 
C_i^{^{3\!}S_1}
\left(1 +
  {\mathcal O}(\chi^2)\right).  \ee
Due to a coupled channel effect,
the same combination of $c_i$ also contributes in the ${^{3\!}}D_1\to {^{3\!}}S_1\to
{^{1\!}}S_0p$ partial wave.  The situation is different when $D$-waves
contribute at the level of the transition operator. The combinations
of $c_i$ in the ${^{3\!}}D_1\to {^{1\!}}S_0p$ amplitude for $pn\to pp\pi^-$
and in the ${^{1\!}}D_2\to {^{3\!}}S_1p$ amplitude for
$pp\to (d/pn) \pi^+$
will influence the
observables, for at the order we are working at there is no contact term
that can absorb the resulting dependence on the LECs $c_i$. It is worth
mentioning that the combinations of $c_i$ in these partial waves are
constrained only weakly by $\pi N$ data. In particular, the
combination of $c_i$ in the ${^{3\!}}D_1$ partial wave, $C_i^{^{3\!}D_1}=c_3-c_4-1/(4M_N)$,
changes from 2 to 7 depending on which of the sets
of $c_i$ given in Ref.~\cite{Evgeny} is used.  Thus, we conclude that the
reaction $NN\to NN \pi$ may serve as an additional source of
information to constrain the $c_i$'s, complementary to the $\pi N$ \cite{Evgeny,bernardLEC} and
$NN$~\cite{timmermans} data (see, however, Ref.~\cite{entem} for some criticism). 
However, for a more quantitative study
of the constraints implied by pion production, and by $\pi N$- and
$NN$ scattering data, a more complete and consistent analysis is 
necessary, which we postpone to a future work.

\section{Summary and Outlook}

We performed a calculation of $p$-wave pion production amplitudes in
$NN$ collisions in three different channels ($pn\to
pp\pi^-$, $pp\to d\pi^+$ and $pp\to pn\pi^+$) in the framework of
chiral effective field theory. The relevant partial
wave transition that depends on the $(N \bar N)^2 \pi$ low-energy constant 
$d$ is ${^{3\!}}S_1\to {^{1\!}}S_0p$ for the first channel and 
${^{1\!}}S_0\to {^{3\!}}S_1p$ for the others. Therefore, it
is clear that the study of different
channels of the pion production reaction $NN \to NN\pi$ probes the 
corresponding operator in very different  kinematical regimes and, thus, 
provides a non-trivial test for the validity of the employed approach. 
Our analysis of all the three channels resulted in values for the LEC $d$ that
are consistent with each other.
In addition, we also point out an inconsistency in the partial wave analysis 
for $pp\to pn\pi^+$ carried out in Ref.~\cite{Flammang}.
Our findings can be interpreted as an indication that the source of the
discrepancy reported in Ref.~\cite{nakamura} is not due to the difference in
the kinematics between $NN \to NN\pi$ and tritium $\beta$-decay, but
rather caused by the inconsistency in the partial wave analysis for 
$pp\to pn\pi^+$ as well as by some technicalities with respect to the work 
of Ref.~\cite{nakamura} that we also discussed in our paper. 
 
Our investigation implies that calculations within effective field theory
yield reliable results for pion production in $NN$ collisions utilizing 
the same value for $d$ even though the corresponding contact term enters 
at very different kinematics in the reactions $pn\to
pp\pi^-$, $pp\to d\pi^+$ and $pp\to pn\pi^+$.
To confirm this conjecture, (i) one needs to reanalyse the reaction 
$pp\to pn\pi^+$ using the complete experimental information available for
$pp\to pp\pi^0$ as input and 
(ii) one needs new data for the process $pn\to pp\pi^-$ at lower
energies. We would like to stress that the near-threshold measurement of the 
reaction $pn\to (pp)_{^{1\!}S_0}\pi^-$, where $(pp)_{^{1\!}S_0}$ signifies that 
the final $pp$ state is constrained to be in the $S$ wave by a kinematical cut,
is the cleanest way to extract information on the contact term from pion
production processes. Such measurements are already under way at COSY. 
Indeed, in the near future both $pp\to
(pp)_{^{1\!}S_0}\pi^0$ and $pn\to (pp)_{^{1\!}S_0}\pi^-$ will be measured
even with polarized initial state~\cite{ANKE}.
In addition, a consistent calculation for both tritium beta decay as
well as low-energy $pd$ scattering should be performed. We plan
to perform these calculations in the future. 

\section*{Acknowledgments}

We would like to thank S. Nakamura for  useful discussions related to the results of his work. 
 The work of E.E. and V.B. was supported in parts by funds provided
from the Helmholtz Association to the young investigator group
``Few-Nucleon Systems in Chiral Effective Field Theory'' (grant
VH-NG-222). This research is part of the EU HadronPhysics2 project ``Study of
strongly interacting matter'' under the Seventh Framework Programme of EU
(Grant agreement n. 227431). Work supported in part by DFG (SFB/TR
16, ``Subnuclear Structure of Matter''), by the 
DFG-RFBR grant (436 RUS 113/991/0-1) and by the Helmholtz
Association through funds provided to the virtual institute ``Spin
and strong QCD'' (VH-VI-231). V.~B., A.~K. and V.~L. acknowledge the support of the
Federal Agency of Atomic Research of the Russian Federation.
V.~L. and A.~K. acknowledge the hospitality of the Institute f\"ur Kernphysik
at FZ J\"ulich.

\appendix

\section{Reaction amplitudes}
\label{amplitudes}
In this Appendix we present expressions for the matrix elements for the reactions we consider.
To calculate them, we used the technique developed in Ref.~\cite{Lensky:Thesis}.

\subsection{General Considerations}
Let us consider pionic reactions involving the $NN$ system, for example 
$NN\to NN\pi$, $\pi d \to NN$, etc. 
In the most general case, an amplitude corresponding to the matrix element
of a particular production and/or absorption operator
between states with given initial ($j,l,s$) and final ($j',l',s'$) total
angular momentum of a 
nucleon pair, its orbital momentum and total spin\footnote{In order to
unambiguously specify the partial wave, 
the pion angular momentum should, in general, also be given. We, however, omit it
since it is only the $p$-wave pion production 
that is  considered here.}
is written as
\begin{eqnarray}
\!\! \!\! \mathcal{A}^\mathrm{full}[jls,j'l's']\!\!  &=&\!\! \maA^\mathrm{tree}[jls,j'l's']\!+\!\maA^\mathrm{FSI}[jls,j'l's']
\!+\!\maA^\mathrm{ISI}[jls,j'l's']
\!+\!\maA^\mathrm{ISI+FSI}[jls,j'l's']\,,
\label{suma}
\end{eqnarray}
where ``tree'' stands for the tree production amplitude, i.e.~where there is no $NN$ (or $N\Delta$)
interaction both in the initial and in the final state, and FSI, ISI, ISI+FSI
refer to the amplitudes with final state, initial state, and both final and
initial state interaction included, in order. 
In this equation we imply that the spin-angular part (as well as the isospin
part) of the amplitudes are factored out. 
Note that since there is a third particle that carries angular
momentum, the pion, the total angular momentum $j$ of the initial two-nucleon
state can be different 
from that of the final two-nucleon state, $j'$. Obviously, the total angular
momentum of the final 
particles has to be equal to that of the initial ones.
Given the tree amplitude as a function of the initial $p$ and final $p'$
relative momenta, $\maA^\mathrm{tree}[jls,j'l's'](p,p')$, 
the remaining amplitudes are given by the following formulae:
\begin{eqnarray}
\maA^\mathrm{FSI}[jls,j'l's']\!\!&=& \!\!
\sum_{l''\!,s''}\int\frac{d^3q}{(2\pi)^3}\frac{\maA^\mathrm{tree}[jls,j'l''s''](p,q)\ 
\maM[j',l''s'',ls](q,p')}%
{4M_{1'}M_{2'}\big[{q^2}/{(2\mu_{1'2'})}-E^\prime-i0\big]}\,,\label{fsi}\\
\maA^\mathrm{ISI}[jls,j'l's']\!\!&=&\!\!
\sum_{l''\!,s''}\int\frac{d^3q}{(2\pi)^3}
\frac{\maM[j,ls,l''s''](p,q)\ \maA^\mathrm{tree}[jl''s'',j'l's'](q,p')}%
{4M_1M_2\big[{q^2}/(2\mu_{12})-E-i0\big]}\,,\label{isi}\\
\maA^\mathrm{ISI+FSI}[jls,j'l's']\!\!&=&\!\!
\sum_{l''\!,s''}\sum_{l'''\!,s'''}
\int\frac{d^3q}{(2\pi)^3}\frac{d^3\ell}{(2\pi)^3}\nonumber\\
& &\hspace{-0.1\textwidth}\times\frac{\maM[j,ls,l''s''](p,q)\
  \maA^\mathrm{tree}[jl''s'',j'l'''s'''](q,\ell)\
  \maM[j',l'''s''',l's'](\ell,p')}%
{4M_1M_2\big[{q^2}/(2\mu_{12})-E-i0\big]\cdot4M_{1'}M_{2'}
  \big[\, {\ell^2}/({2\mu_{1'2'}})-E^\prime-i0\big]}\,,\label{fsiisi}
\end{eqnarray}
where $M_{1,2}$ ($M_{1',2'}$) are the masses of the particles in the
intermediate state that are related via the NN interaction  to the initial (final) state, $\mu_{12}$
($\mu_{1'2'}$) are the corresponding reduced masses, $E$ ($E^\prime$) is the
energy of the initial (final) two-nucleon state in its center-of-mass frame,
$\maM[j,l_i s_i,l_f s_f]$ is the $NN$ half-offshell $\maM$-matrix
corresponding to a transition from the state $(jl_i s_i)$ to the state $(jl_f
s_f)$, and the sums are over all the intermediate states with given $j,\ j'$,
$l$, $l^\prime$, $s$, and $s^\prime$.  We use the following relation between the
$\maM$-matrix and the commonly used $\mathcal T$-matrix:
$\maM=-8\pi^2\sqrt{M_1M_2M_3M_4}~\mathcal T$, where the $M_i$ are the masses 
of interacting particles. 

The formulae given above also hold for the case when there is a transition
through an intermediate 
$N\Delta$ state going to a final (from an initial) state via an $NN-N\Delta$
interaction. In this case the $NN$ $\maM$-matrices have to be replaced by the 
appropriate $NN-N\Delta$ matrices, and also the 
propagators entering Eqs.~(\ref{fsi})-(\ref{fsiisi}) that correspond to the
$N\Delta$ intermediate state have to be modified       
according to
\begin{eqnarray}
\frac{1}{4M_1M_2\big[{q^2}/(2\mu_{12})-E-i0\big]}& {\longrightarrow}
&\frac{1}{4\sqrt{2}M_1M_2\big[{q^2}/(2\mu_{12})-(E-\Delta M)-i0\big]}\,, 
\end{eqnarray}
where $\Delta M$ is the nucleon-$\Delta$ mass difference (note also the factor
$1/\sqrt{2}$). Of course, a tree diagram with a $N\Delta$ initial or final
state gives a nonzero contribution only when it is inserted as a building block into
those of FSI and ISI diagrams that have $N\Delta$ as an intermediate state. 

In case of a deuteron in the final state, the corresponding $\mathcal M$ matrices
should be replaced by the deuteron wave functions according to
\begin{eqnarray}
\maA^\mathrm{FSI}[jls,1]&=& 
\frac{1}{\sqrt{2M_N}}
\sum_{l^{\prime\prime}}\int\frac{d^3q}{(2\pi)^3}\ \maA^\mathrm{tree}[jls,1l''s''](p,q)
i^{l''}\psi^{l''}(q),
\label{dfsi}
\end{eqnarray}
where $\psi^{l^{\prime\prime}}(q)$ are the deuteron wave functions corresponding 
to the angular momentum $l^{\prime\prime}$, normalized by the condition 
\beq
\int \frac{d^3 q}{(2\pi)^3}\left( (\psi^0(q))^2+ (\psi^2(q))^2\right)=1.
\eeq 
Thus, the two-nucleon propagator for 
the deuteron in the final state is absorbed in the wave functions and the normalization 
has changed. 
Analogous expressions can be written down for the deuteron in the initial state
and also for the deuteron in the initial and final states. 
Note that in the case of the deuteron in the inital and/or final state
the tree diagrams appear only as building blocks for the calculation of the
ISI/FSI and ISI+FSI diagrams according to Eqs.~(\ref{fsi})-(\ref{fsiisi}) and \ref{dfsi}),
respectively. They do not contribute independently because then there are no free 
nucleons in the initial and/or final state.

\subsection{The reaction {\boldmath$pn\to pp\pi^-$}}
Here and below we use the spectroscopic notation $^{2S+1}L_J$ for the $NN$ and $N\Delta$
partial waves rather than the $[jls]$ notation used in the previous section.
The transitions that contribute to the reaction $pn\to pp\pi^-$ at energies close to threshold
are $^{3\!}S_1\to{^1\!}S_0p,\ ^{3\!}D_1\to {^1\!S_0p}$ in the isospin-zero
initial state, and $^3\!P_0\to {^1\!}S_0s$ in the isospin-one initial
state. The spin-angular structure of the amplitude reads 
\begin{equation}
\maM_{pn\to pp\pi^-} =\bigg[A_1(\vec \mathcal{S}\hat p)+C_1(\vec \mathcal{S}\hat k_\pi)+C_2\,
\vec \mathcal{S}((\hat p\hat k_\pi)\hat p-\frac{1}{3}\hat k_\pi)\bigg]
\mathcal{I}^{\prime\dagger}, 
\label{amppimin}
\end{equation}
where $\vec \mathcal{S}=\chi^T_2\frac{\sigma_2}{\sqrt{2}}\vec \sigma\chi_1$,
$\mathcal{I}^\prime=\chi^T_{2^\prime}\frac{\sigma_2}{\sqrt{2}}\chi_{1^\prime}$
denote normalized spin structures corresponding to the initial spin-triplet and
final spin-singlet states, in order. 
Here and below, $\hat p,\hat p\,',\hat k_\pi$ denote 
unit vectors of initial and final relative momenta of two nucleons and that of
the pion momentum, respectively, and the $\chi$'s with corresponding indices stand
for the spinors of the initial and final nucleons.
In turn, $A_1$, $C_1$, and $C_2$ are the amplitudes corresponding to the
$^3P_0\to {^1\!}S_0s$, ${^3\!}S_1\to{^1\!}S_0p$, and 
$^{3\!}D_1\to{^1\!}S_0p$ transitions, in order. They are related to 
the corresponding amplitudes in the $JLS$ basis via
\begin{eqnarray}
A_1&=&\frac{1}{\sqrt{3}}\ \maA^\mathrm{full}[^3\!P_0,{^1\!}S_0],\label{a1eq}\\
C_1&=&\maA^\mathrm{full}[^{3\!}S_1,{^1\!}S_0],\\
C_2&=&\frac{3}{\sqrt{2}}\ \maA^\mathrm{full}[^{3\!}D_1,{^1\!S_0}].
\end{eqnarray}
The observables we consider are expressed in terms of these amplitudes in the following way:
\begin{eqnarray}
 \frac{d^2\sigma}{d\Omega d m^2_{pp}} &=&\frac{1}{4\Delta M_\mathrm{pp}^2}
\frac{1}{(4\pi)^4M_N s
  p}\int\limits_0^{p_\mathrm{cut}}k_\pi(p')p'^2dp'\Bigg[|A_1|^2+\bigg|C_1-\frac{C_2}
{3}\bigg|^2\nonumber\\&&+2\mathrm{Re}\, 
\big(A_1^*(C_1+\frac{2C_2}{3})\big)\cos\theta_{\pi}+\bigg[2\mathrm{Re}\,(C_2^*C_1)+
\frac{|C_2|^2}{3}\bigg]\cos^2\theta_{\pi}\Bigg]\,,\\   
A_y\cdot\frac{d^2\sigma}{d\Omega d m^2_{pp}}&=&\frac{1}{4\Delta M_\mathrm{pp}^2}
\frac{1}{(4\pi)^4M_N s p}\int\limits_0^{p_\mathrm{cut}}k_\pi(p')p'^2dp'
\nonumber\\
& & \times \bigg[\sin 2\theta_{\pi}\,\mathrm{Im}\, (C_1^*C_2)
-2\sin\theta_{\pi}\,\mathrm{Im}\,(A_1^*(C_1-\frac{C_2}{3}))\bigg]\, ,
\end{eqnarray}
where $p_\mathrm{cut}$ is the maximum relative momentum of the final protons in the measurements at TRIUMF~\cite{hahn,duncan},
$\Delta M_\mathrm{pp}^2=(2M_N+p^2_{cut}/M_N)^2-(2M_N)^2\approx 4p_\mathrm{cut}^2$, 
$k_\pi(p')$ is the momentum of the final pion, and $s$ and $p$ are the invariant 
energy squared and the relative momentum of the initial 
nucleons, in order.

Below we give the expressions for the tree amplitudes
$\maA[^{3\!}S_1,{^1\!}S_0]$, $\maA[^{3\!}D_1,{^1\!S_0}]$ resulting from
various pion production mechanisms as well as those for the relevant 
production amplitudes involving the $\Delta$ isobar. 
Note that from here on we suppress the label "tree" on the tree-level transition amplitudes.
We also explain how we extract $A_1$ from experimental data.

\subsubsection{Direct production}
\begin{eqnarray}
\!\!\!\!\maA[^{3\!}S_1,{^1\!}S_0](p,p')&=&C\int\frac{d\Omega_{\vec
    k_\pi}}{4\pi}\bigg[\!\!-\!k_\pi\!+\!\frac{\omega_\pi}{M_N}(\vec p\,'\hat
k_\pi)\bigg](2\pi)^3\delta^{(3)}(\vec p\,'\!-\!\vec p\!+\!\vec k_\pi/2)\,,\\ 
\!\!\!\!\maA[^{3\!}D_1,{^1\!S_0}](p,p')& =
&\frac{C}{\sqrt{2}}\int\frac{d\Omega_{\vec
    k_\pi}}{4\pi}\bigg[\!\!-\!k_\pi[3(\hat p\hat
k_\pi)^2\!-\!1]\!+\!\frac{\omega_\pi}{M_N}[3(\hat p\hat k_\pi)(\hat p\vec
p\,')\!-\!(\vec p\,'\hat k_\pi)]\bigg] 
\nonumber\\
& &\times
(2\pi)^3\delta^{(3)}(\vec p\,'\!-\!\vec p\!+\!\vec k_\pi/2)\,,
\end{eqnarray}
where $C=-\mathrm{i}\displaystyle\frac{8M_N^2g_A}{f_\pi\sqrt{2}}$, and
$\omega_\pi=\sqrt{k_\pi^2+m_\pi^2}$ is the 
energy of the final pion. Here we included both the leading $\pi NN$ vertex
and its recoil correction which 
enters at NNLO.

\subsubsection{Production via the {$\Delta(1232)$} isobar}
The $\Delta(1232)$ contribution comes from the $N\Delta$ intermediate states.
In the reaction $pn\to pp\pi^-$ with the initial isospin of the $pn$ system being $I=0$,
the $N\Delta\leftrightarrow NN$ transitions are allowed only in the final
state interaction. As we consider those 
kinematical configurations where the relative kinetic energy of the final protons is small,
it is only the ${^1\!}S_0$ final state that contributes. Therefore, the only coupled channel
where the $\Delta(1232)$ contributes is ${^5\!}D_0(N\Delta)\to {^1\!S_0}(NN)$. For $p$ wave pions 
the relevant amplitudes that correspond to the ${^3\!}S_1(NN)\to{^5\!}D_0(N\Delta)$ and 
${^3\!}D_1(NN)\to{^5\!}D_0(N\Delta)$ transitions
in the production operator read:
\begin{eqnarray}
\!\!\!\!\maA[{^3\!}S_1,{^5\!}D_0](p,p')&=&C_\Delta\int\frac{d\Omega_{\vec
    k_\pi}}{4\pi}\bigg[\!\!-\!k_\pi[3(\hat p\,'\hat
k_\pi)^2\!-\!1]\bigg](2\pi)^3\delta^{(3)}(\vec p\,'\!-\!\vec p\!+\!\vec k_\pi
\vartheta)\,,\\ 
\!\!\!\!\maA[{^3\!}D_1,{^5\!}D_0](p,p')& =
&\frac{C_\Delta}{\sqrt{2}}\int\frac{d\Omega_{\vec
    k_\pi}}{4\pi}\bigg[\!\!-\!k_\pi[9(\hat p\,'\hat  p)(\hat p\,'\hat
k_\pi)(\hat p\hat k_\pi)\!-\!3(\hat p\hat k_\pi)^2\!-\!3(\hat p\,'\hat
k_\pi)^2\!+\!1]\bigg]\nonumber\\ 
& &\times(2\pi)^3\delta^{(3)}(\vec p\,'\!-\!\vec p\!+\!\vec k_\pi\vartheta)\,,
\end{eqnarray}
where $C_\Delta=-\mathrm{i}\displaystyle\frac{8M_NM_\Delta
  h_A}{3f_\pi\sqrt{2}}\sqrt{\frac{M_N}{M_\Delta}}$, and
$\vartheta=\displaystyle\frac{M_N}{M_N+M_\Delta}$ and $p'$ is the relative momentum of the $N\Delta$ state. 

\subsubsection{Rescattering via the {$s$}-wave WT vertex}
 
\begin{eqnarray}
\!\!\!\!\maA[{^3\!}S_1,{^1\!}S_0](p,p')&=& -\frac{C\,\omega_\pi
  k_\pi}{2f_\pi^2}\int\frac{d\Omega_{\vec p}}{4\pi} 
 \frac{1}{(\vec p-\vec p\,')^2+m_\pi^2} 
\left[1-\frac{2(\vec p-\vec p\,')^2}{3[(\vec p-\vec p\,')^2+m_\pi^2]}\right]\,,\\
\!\!\!\!\maA[{^3\!}D_1,{^1\!}S_0](p,p')&=& \frac{C\,\omega_\pi
  k_\pi}{3f_\pi^2\sqrt{2}}\int\frac{d\Omega_{\vec p}}{4\pi} 
\frac{3(\vec p\,'\hat p-p)^2-(\vec p-\vec p\,')^2}{[(\vec p-\vec p\,')^2+m_\pi^2]^2}\,.
\end{eqnarray}

Note that in these expressions, and also in the expressions 
for the amplitudes that stem from operators with $c_3$, $c_4$, 
and recoil corrections to the WT vertex (see below), we keep only 
the leading term in the expansion in powers of $k_\pi/p$. The same 
is true for the corresponding amplitudes in the reactions $pp\to d\pi^+$ and $pp\to pn\pi^+$.

\subsubsection{Operators with {$c_3$, $c_4$}, and  recoil  corrections to the WT vertex}
\begin{eqnarray}
 \!\!\!\!\maA[{^3\!}S_1,{^1\!}S_0](p,p')&=&\frac{4C\,k_\pi}{3f_\pi^2}\int\frac{d\Omega_{\vec p\,'}}{4\pi}\left[
C^{^3\!S_1}_i\frac{(\vec p-\vec p\,')^2}{(\vec p-\vec p\,')^2+m_\pi^2}
+\frac{1}{8M_N}\frac{p'^2-p^2}{(\vec p-\vec p\,')^2+m_\pi^2}
\right]\,, \\
\!\!\!\!\maA[{^3\!}D_1,{^1\!}S_0](p,p')&=&\frac{2C\,k_\pi}{3f_\pi^2\sqrt{2}}\int\frac{d\Omega_{\vec
    p\,'}}{4\pi}\left[ 
C^{^3\!D_1}_i\frac{3(\vec p\,'\hat p-p)^2-(\vec p-\vec p\,')^2}{(\vec p-\vec p\,')^2+m_\pi^2}
\right.
\nonumber\\
& &\left.
+\frac{1}{4M_N}\frac{3((\vec p\,'\hat p)^2-p^2)-p'^2+p^2}{(\vec p-\vec p\,')^2+m_\pi^2}
\right],
\end{eqnarray}
where
$C^{^3\!S_1}_i=\displaystyle\frac{c_3}{2}+c_4+\displaystyle\frac{1}{4M_N}$,
$C^{^3\!D_1}_i=c_3-c_4-\displaystyle\frac{1}{4M_N}$. 

\subsubsection{Contact term}
\begin{eqnarray}
 \!\!\!\!\maA[{^3\!}S_1,{^1\!}S_0](p,p')&=&\frac{2C\,k_\pi}{g_A}d\,,\\
\!\!\!\!\maA[{^3\!}D_1,{^1\!}S_0](p,p')&=&\ 0.
\end{eqnarray}

\subsubsection{Coulomb interaction}
Since the final two protons are at low relative momenta, there are
sizable effects from the Coulomb interaction
between the two protons.  The effect of the Coulomb interaction was
taken into account along the lines of
Refs.~\cite{Hanhart:1995ut,Baru:2000hg}. To be specific, we multiply all tree
diagrams that do not contain the $\Delta$ isobar by the Gamow-Sommerfeld factor
\begin{eqnarray}
G(p')&=&\bigg[\frac{2\pi\gamma(p')}{\exp 2\pi\gamma(p')-1}\bigg]^{1/2},\\
\gamma(p')&=&\frac{M_N}{2\alpha p'},
\nonumber
\end{eqnarray}
where $\alpha$ is the electromagnetic fine structure constant. At the same
time, the half-offshell $pp$ $\maM$~matrix 
in the ${^1\!}S_0$ partial wave that we use in our calculation is corrected for
the Coulomb interaction according to 
\begin{eqnarray}
\maM^{\mathrm{CC}}(q,p')&=&\maM(q,p')\frac{G(q)}{G(p')}\frac{\maM^{\mathrm{CC}}(p',p')}{\maM(p',p')},
\end{eqnarray}
where $\maM(q,p')$ and $\maM^{\mathrm{CC}}(q,p')$ are, in order,
the half-offshell $pp$ $\maM$~matrices 
without and with the inclusion of the Coulomb interaction, whereas $\maM(p',p')$ and
$\maM^{\mathrm{CC}}(p',p')$ are  the 
corresponding $pp$ onshell $\maM$~matrices.
See Ref.~\cite{Hanhart:1995ut} for more details.

As far as diagrams with the $\Delta$ are concerned, where we have the transition
${^5\!}D_0(N\Delta)\to {^1\!S_0}(NN)$, 
we apply the following argument in order to take into account the Coulomb
interaction. First, we note that the 
typical relative momenta of the intermediate $N\Delta$ state are large so that
the Coulomb interaction in this intermediate state is expected to be
unimportant. In order to take into account the Coulomb interaction
between the protons in the final state, we multiply the amplitude with
$N\Delta$ by the ratio of the inverse Jost functions
\begin{eqnarray}
G_{N\Delta}(p')&=&\frac{F_\mathrm{pp}^\mathrm{CC}(p')}{F_\mathrm{pp}(p')},
\end{eqnarray}
where $F_\mathrm{pp}^\mathrm{CC}(p')$ and $F_\mathrm{pp}(p')$ are the $pp$ inverse Jost functions with
and without Coulomb interaction, respectively. They are related to the
corresponding $\maM$~matrices via
\begin{eqnarray}
F_\mathrm{pp}^\mathrm{CC}(p')&=&G(p')+\frac{1}{4M_N}\int\frac{d^3q}{(2\pi)^3}
\frac{G(q)\maM^{\mathrm{CC}}(q,p')}{q^2-p'^2-i0}\,,\\  
F_\mathrm{pp}(p')&=&1+\frac{1}{4M_N}\int\frac{d^3q}{(2\pi)^3}
\frac{\maM(q,p')}{q^2-p'^2-i0}. 
\end{eqnarray}
Notice that the overall normalization of the Jost functions  is of
no relevance, for they only occur in the ratio 
(however, both Jost functions have to have the same normalization factors).

\subsubsection{The contribution of the {$I=1$} initial state}

We write $A_1$ in Eq.~(\ref{amppimin}) in a form which takes into account the 
ISI phase as well as the dependence of the final state Jost function on the momenta:
\begin{eqnarray}
A_1&=&\mathrm{i}\,X\exp(\mathrm{i}\delta^{{^3\!}P_0})F_\mathrm{pp}^\mathrm{CC}(p').
\end{eqnarray}
The constant (real) factor $X$ is adjusted in such a way to reproduce the corresponding amplitude
extracted from the TRIUMF data using a partial wave analysis~\cite{duncan,duncan2}.
The sign of $X$ is adjusted 
to the behaviour of observables in $pn\to pp\pi^-$.

\subsection{The reaction {\boldmath$pp\to d\pi^+$}}
\label{reacdpi}

The transitions that contribute to the reaction $pp\to d\pi^+$ at energies close to threshold
are ${^1\!}S_0\to{^3\!}S_1p,\ {^1\!}S_0\to{^3\!}D_1p$,
${^1\!}D_2\to{^3\!}S_1p,\ {^1\!}D_2\to{^3\!}D_1p$, and ${^3\!}P_1\to
{^3\!}S_1s$, ${^3\!}P_1\to {^3\!}D_1s$. The spin-angular structure of the
amplitude reads 
\begin{equation}
\maM_{pp\to d\pi^+}=\bigg[C_0\,(\vec\mathcal{S}\times \hat
p\,)\vec\varepsilon\,+C_1\mathcal{I}(\hat k\vec\varepsilon\,)+C_2\,\mathcal{I}
[(\hat p\hat k)(\hat p\vec\varepsilon\,)-\frac{1}{3}(\hat
k\vec\varepsilon\,)]\bigg], 
\end{equation}
where $\vec\varepsilon$ is the deuteron polarization vector, $\vec
\mathcal{S}=\chi^T_2\frac{\sigma_2}{\sqrt{2}}\vec \sigma\chi_1$,
$\mathcal{I}=\chi^T_{2}\frac{\sigma_2}{\sqrt{2}}\chi_{1}$ are normalised spin
structures corresponding to the initial spin-triplet and spin-singlet states,
in order. Here, the $\chi$'s refer to spinors of the initial nucleons, 
and $C_0$, $C_1$, and $C_2$ are the amplitudes corresponding to the
${^3\!}P_1\to \vec\varepsilon\, s$, ${^1\!}S_0\to \vec\varepsilon\, p$, and 
${^1\!}D_2\to \vec\varepsilon\, p$ transitions, in order. Further, we denote the
final states as $\vec \varepsilon\, l$, where 
$\vec\varepsilon$ stands for the deuteron final state, and $l$ is the angular
momentum of the final pion relative to the deuteron.  
The amplitudes  $C_0$, $C_1$, and $C_2$ are related to the corresponding
amplitudes in $JLS$ basis via 
\begin{eqnarray}
C_0&=&\sqrt{\frac{3}{2}}\ \maA^\mathrm{full}[{^3\!}P_1,1]\,,\label{c0r}\\
C_1&=& \maA^\mathrm{full}[{^1\!}S_0,1]\,,\label{c1r}\\
C_2&=&\sqrt{\frac{15}{2}}\ \maA^\mathrm{full}[{^1\!}D_2,1].\label{c2r}
\end{eqnarray}
Note that, as it can be seen from Eq.~(\ref{dfsi}), the amplitudes of
the transitions to the deuteron state are 
sums of the amplitudes where the transition goes to the $S$-wave 
component and those where the transition goes to the $D$-wave component
of the deuteron wave function. Note also that
here the total amplitudes are distinguished 
only by the initial state, as the final state, the deuteron, is the same for
all transitions. However, at the level of tree amplitudes, one has to
distinguish between the ones that correspond to transitions to the $S$-wave
component and those to the $D$-wave component.

The observables under consideration are expressed through the amplitudes $C_0$,
$C_1$, and $C_2$ as 
\begin{eqnarray}
\frac{d\sigma}{d\Omega}&=& \frac{k_\pi}{256\pi^2
  sp}\bigg[2|C_0|^2+|C_1|^2+\frac{1}{9}|C_2|^2(3\cos^2\theta_{\pi}+1)+\frac{2}{3}\mathrm{Re}\,
(C_1C_2^*)(3\cos^2\theta_{\pi}-1)\bigg]\,, 
\nonumber\\ & &\\
A_y\cdot\frac{d\sigma}{d\Omega}&=&\frac{k_\pi}{256\pi^2 sp}\cdot
2\sin\theta_{\pi}\cos\phi\ \mathrm{Im}\,\big(C_0^*(C_1-\frac{C_2}{3})\big). 
\end{eqnarray}

Below, we give expressions for the tree amplitudes
$\maA[{^1\!}D_2,{^3\!}D_1]$, $\maA[{^1\!}D_2,{^3\!}S_1]$,
$\maA[{^1\!}S_0,{^3\!}D_1]$, $\maA[{^1\!}S_0,{^3\!}S_1]$, contributing to
$C_1$ and $C_2$, 
as well as  those for the relevant 
production amplitudes involving the $\Delta$ isobar. 
We also provide details of the determination of $C_0$.

\subsubsection{Direct production}

\begin{eqnarray}
\!\!\!\! \maA[{^1\!}S_0,{^3\!}S_1](p,p')&=& C\int\frac{d\Omega_{\vec
    k_\pi}}{4\pi}\bigg[\!\!-\!k_\pi\!+\!\frac{\omega_\pi}{M_N}(\vec p\,'\hat
k_\pi)\bigg](2\pi)^3\delta^{(3)}(\vec p\,'\!-\!\vec p\!+\!\vec k_\pi/2)\,,\\ 
\!\!\!\!
\maA[{^1\!}S_0,{^3\!}D_1](p,p')&=&\frac{C}{\sqrt{2}}\int\frac{d\Omega_{\vec
    k_\pi}}{4\pi}\bigg[\!\!-\!k_\pi[3(\hat p\,'\hat
k_\pi)^2\!-\!1]\!+\!\frac{2\omega_\pi}{M_N} (\vec p\,'\hat k_\pi)\bigg] 
\nonumber\\
& &
\times(2\pi)^3\delta^{(3)}(\vec p\,'\!-\!\vec p\!+\!\vec k_\pi/2)\,, \\
\!\!\!\!
\maA[{^1\!}D_2,{^3\!}S_1](p,p')&=&C\sqrt{\frac{3}{10}}\int\frac{d\Omega_{\vec
    k_\pi}}{4\pi} 
\bigg[\!\!-\!k_\pi[3(\hat p\hat
k_\pi)^2\!-\!1]\!+\!\frac{\omega_\pi}{M_N}[3(\hat p\vec p\,')(\hat p\hat
k_\pi)\!-\!\vec p\,'\hat k_\pi]\bigg] 
\nonumber\\
& &
\times(2\pi)^3\delta^{(3)}(\vec p\,'\!-\!\vec p\!+\!\vec k_\pi/2)\,, \\
\!\!\!\! \maA[{^1\!}D_2,{^3\!}D_1](p,p')&=&
\frac{C}{\sqrt{2}}\sqrt{\frac{3}{10}}\int\frac{d\Omega_{\vec k_\pi}}{4\pi}
\bigg[\!\!-\!k_\pi[9(\hat p\,'\hat k_\pi)(\hat p\,'\hat p)(\hat p\hat
k_\pi)\!-\!3(\hat p\hat k_\pi)^2\!-\!3(\hat p\,'\hat
k_\pi)^2\!+\!1]\nonumber\\ 
& &+\frac{2\omega_\pi}{M_N}[3(\hat p\vec p\,')(\hat p\hat k_\pi)\!-\!\vec
p\,'\hat k_\pi]\bigg](2\pi)^3\delta^{(3)}(\vec p\,'\!-\!\vec p\!+\!\vec
k_\pi/2)\,. 
\end{eqnarray}

\subsubsection{Production via the {$\Delta(1232)$} isobar}

In the reaction $pp\to d\pi^+$, the initial isospin of the $pp$ system is $I=1$,
so the $N\Delta\leftrightarrow NN$ intermediate states are allowed only in the
initial state interaction. 
The coupled channels that contribute are\footnote{
We do not take into account the channel ${^1\!}D_2(NN)\to{^3\!}D_2(N\Delta)$
because the corresponding $NN\to N\Delta$ $\maM$-matrix is subleading
according to the power counting and it is also
numerically small at the energies considered~\cite{CCF}.}
${^1\!}S_0(NN)\to{^5\!}D_0(N\Delta)$,
${^1\!}D_2(NN)\to{^5\!}D_2(N\Delta)$, and ${^1\!}D_2(NN)\to{^5\!}S_2(N\Delta)$.
The relevant transitions in the production operator for $p$-wave pions are $^5\!D_0(N\Delta)\to {^3\!S_1}(NN)$,
$^5\!D_0(N\Delta)\to {^3\!D_1}(NN)$, $^5\!D_2(N\Delta)\to {^3\!S_1}(NN)$,
$^5\!D_2(N\Delta)\to {^3\!D_1}(NN)$,$^5\!S_2(N\Delta)\to {^3\!S_1}(NN)$, and
$^5\!S_2(N\Delta)\to {^3\!D_1}(NN)$. The expressions for the corresponding
amplitudes read: 
\begin{eqnarray}
\!\!\!\!\maA[{^5\!}D_0,{^3\!}S_1](p,p')&=& C_\Delta\int\frac{d\Omega_{\vec k_\pi}}{4\pi}
\bigg[-k_\pi[3(\hat p\hat k_\pi)^2-1]\bigg] (2\pi)^3\delta^{(3)}(\vec
p\,'\!-\!\vec p\!+\!\vec k_\pi/2)\,,\\ 
\!\!\!\!\maA[{^5\!}D_0,{^3\!}D_1](p,p')&=&
\frac{C_\Delta}{\sqrt{2}}\int\frac{d\Omega_{\vec
    k_\pi}}{4\pi}\bigg[\!\!-\!k_\pi[9(\hat p\,'\hat  p)(\hat p\,'\hat
k_\pi)(\hat p\hat k_\pi)\!-\!3(\hat p\hat k_\pi)^2\!-\!3(\hat p\,'\hat
k_\pi)^2\!+\!1]\bigg] 
\nonumber\\
& &
\times
(2\pi)^3\delta^{(3)}(\vec p\,'\!-\!\vec p\!+\!\vec k_\pi/2)\,,\\
\!\!\!\!\maA[{^5\!}D_2,{^3\!}S_1](p,p')&=&C_\Delta\sqrt{\frac{21}{20}}\int\frac{d\Omega_{\vec
    k_\pi}}{4\pi}\bigg[\!\!-\!k_\pi[3(\hat p\hat
k_\pi)^2\!-\!1]\bigg](2\pi)^3\delta^{(3)}(\vec p\,'\!-\!\vec p\!+\!\vec
k_\pi/2)\,, \\ 
\!\!\!\!\maA[{^5\!}D_2,{^3\!}D_1](p,p')&=&\frac{3}{4}C_\Delta\sqrt{\frac{6}{35}}\int\frac{d\Omega_{\vec
    k_\pi}}{4\pi} 
\bigg[\!\!-\!k_\pi\big[9(\hat p\hat p\,')^2\!-\!6(\hat p\hat p\,')(\hat p\hat
k_\pi) (\hat p\,'\hat k_\pi)\nonumber\\ 
& &+\!2(\hat p\hat k_\pi)^2\!+\!2(\hat p\,'\hat
k_\pi)^2-\frac{11}{3}\big]\bigg](2\pi)^3\delta^{(3)}(\vec p\,'\!-\!\vec
p\!+\!\vec k_\pi/2)\,, \\ 
\!\!\!\!\maA[{^5\!}S_2,{^3\!}S_1](p,p')&=&
C_\Delta\sqrt{6}\int\frac{d\Omega_{\vec k_\pi}}{4\pi}
[-k_\pi] (2\pi)^3\delta^{(3)}(\vec p\,'\!-\!\vec p\!+\!\vec k_\pi/2)\,,\\
\!\!\!\!\maA[{^5\!}S_2,{^3\!}D_1](p,p')&=&
\frac{\sqrt{3}}{10}C_\Delta\int\frac{d\Omega_{\vec k_\pi}}{4\pi}
\bigg[-k_\pi[3(\hat p\,'\hat k_\pi)^2-1]\bigg] (2\pi)^3\delta^{(3)}(\vec p\,'\!-\!\vec p\!+\!\vec k_\pi/2)\,.
\end{eqnarray}

\subsubsection{Rescattering via the {$s$}-wave WT vertex}

\begin{eqnarray}
\!\!\!\!\maA[{^1\!}S_0,{^3\!}S_1](p,p')&=&
\frac{C\,\omega_\pi k_\pi}{2f_\pi^2}\int\frac{d\Omega_{\vec p}}{4\pi}
 \frac{1}{(\vec p-\vec p\,')^2+m_\pi^2}
 \left[1-\frac{2(\vec p-\vec p\,')^2}{3[(\vec p-\vec p\,')^2+m_\pi^2]}\right]\,, \\
\!\!\!\!\maA[{^1\!}S_0,{^3\!}D_1](p,p')&=&
 -\frac{C\,\omega_\pi k_\pi}{3f_\pi^2\sqrt{2}}\int\frac{d\Omega_{\vec p}}{4\pi}
\frac{3(\vec p\hat p'-p')^2-(\vec p-\vec p\,')^2}{[(\vec p-\vec p\,')^2+m_\pi^2]^2}\,,\\
\!\!\!\!\maA[{^1\!}D_2,{^3\!}S_1](p,p')&=&
 -\frac{C\,\omega_\pi k_\pi}{f_\pi^2\sqrt{30}}\int\frac{d\Omega_{\vec{p}^\prime}}{4\pi}\frac{3(\vec p\,'\hat p\!-\!p)^2
 -(\vec p\!-\!\vec p\,')^2}{[(\vec p-\vec p\,')^2+m_\pi^2]^2}\,,\\
\!\!\!\!\maA[{^1\!}D_2,{^3\!}D_1](p,p')&=&
\frac{3}{\sqrt{2}}\frac{C\,\omega_\pi k_\pi}{f_\pi^2\sqrt{30}}\int\frac{d\Omega_{\vec{p}^\prime}}{4\pi}\frac{1}{(\vec p-\vec p\,')^2+m_\pi^2}\bigg[\frac{3(\hat p\hat p')^2-1}{2}\nonumber\\
& -&\frac{9(p\!-\!\vec p\,'\hat p)(\vec p\hat p'\!-\!p')(\hat p\hat p')\!-\!3(p\!-\!\vec p\,'\hat p)^2\!-\!3(p'\!-\!\vec p\hat p'\,)^2\!+\!(\vec p\!-\!\vec p\,')^2}{3[(\vec p-\vec p\,')^2+m_\pi^2]}\bigg]\,.
\end{eqnarray}

\subsubsection{Operators with {$c_3$, $c_4$}, and recoil  corrections to the WT vertex}

\begin{eqnarray}
\!\!\!\!\maA[{^1\!}S_0,{^3\!}S_1](p,p')&=&
\frac{4C\,k_\pi}{3f_\pi^2}\int\frac{d\Omega_{\vec{p}^\prime}}{4\pi}\left[
C^{^3\!S_1}_i\frac{(\vec p-\vec p\,')^2}{(\vec p-\vec p\,')^2+m_\pi^2}
+\frac{1}{8M_N}\frac{p^2-p'^2}{(\vec p-\vec p\,')^2+m_\pi^2}
\right]\,, \\
\!\!\!\!\maA[{^1\!}S_0,{^3\!}D_1](p,p')&=&
\frac{2C\,k_\pi}{3f_\pi^2\sqrt{2}}
\int\frac{d\Omega_{\vec{p}^\prime}}{4\pi}\left[
C^{^3\!D_1}_i\frac{3(\vec p\hat p'-p')^2-(\vec p-\vec p\,')^2}{(\vec p-\vec p\,')^2+m_\pi^2}
\right.
\nonumber\\
& &
\left.
+\frac{1}{4M_N}\frac{3((\vec p\hat p')^2-p'^2)-p^2+p'^2}{(\vec p-\vec p\,')^2+m_\pi^2}
\right]\,,\\
\!\!\!\!\maA[{^1\!}D_2,{^3\!}S_1](p,p')&=&
\frac{2C\,k_\pi}{f_\pi^2\sqrt{30}}
\int\frac{d\Omega_{\vec{p}^\prime}}{4\pi}\left[
C^{^3\!D_1}_i\frac{3(\vec p\,'\hat p-p)^2-(\vec p-\vec p\,')^2}{(\vec p-\vec p\,')^2+m_\pi^2}
\right.
\nonumber\\
& &
\left.
-\frac{1}{4M_N}\frac{3((\vec p\,'\hat p)^2-p^2)-p\,'^2+p^2}{(\vec p-\vec p\,')^2+m_\pi^2}
\right]\,,\\
\!\!\!\!\maA[{^1\!}D_2,{^3\!}D_1](p,p')&=&
\frac{C\,k_\pi}{f_\pi^2\sqrt{15}}
\int\frac{d\Omega_{\vec{p}^\prime}}{4\pi}
\left[
C^{^3\!D_1}_i
\left\{
\frac{9(p\!-\!\vec p\,'\hat p)(\vec p\hat p'\!-\!p')(\hat p\hat p')}{(\vec p-\vec p\,')^2+m_\pi^2}
\right.
\right.
\nonumber\\
& &
\left.
+\frac{-\!3(p\!-\!\vec p\,'\hat p)^2\!-\!3(p'\!-\!\vec p\hat p'\,)^2\!+\!(\vec
  p\!-\!\vec p\,')^2}{(\vec p-\vec p\,')^2+m_\pi^2} 
\right\}
\nonumber\\
& &
-\frac{1}{4M_N}
\left\{
\frac{9(\vec p\,'\hat p+\!p)( p\,'\!\!-\!\vec p\hat p')(\hat p\hat
  p')}{(\vec p-\vec p\,')^2+m_\pi^2} 
\right.
\nonumber\\
& &
\left.
+\frac{-\!3(p'\!+\!\vec p\hat p')(p'\!-\!\vec p\hat p\,')\!-\!3(\vec p\,'\hat
  p+p)(\vec p\,'\hat p-p)+p'^2\!-\!p^2}{(\vec p-\vec p\,')^2+m_\pi^2} 
\right\}
\nonumber\\
& &\left.+3\left(c_4+\frac{1}{4M_N}\right)\frac{[3(\hat p\hat p')^2-1](\vec
    p\!-\!\vec p\,')^2}{(\vec p-\vec p\,')^2+m_\pi^2} 
\right]\,.
\end{eqnarray}

\subsubsection{Contact term}

\begin{eqnarray}
\!\!\!\!\maA[{^1\!}S_0,{^3\!}S_1](p,p')&=&\frac{2C\,k_\pi}{g_A}d\,,\\
\!\!\!\!\maA[{^1\!}S_0,{^3\!}D_1](p,p')&=&\ 0\,,\\
\!\!\!\!\maA[{^1\!}D_2,{^3\!}S_1](p,p')&=&\ 0\,, \\
\!\!\!\!\maA[{^1\!}D_2,{^3\!}D_1](p,p')&=&\ 0\,.
\end{eqnarray}

\subsubsection{Coulomb interaction}
For this reaction, the Coulomb interaction between the initial protons gives 
only a small effect since the initial 
relative momentum is large. On the contrary, the deuteron and pion in the
final state are at low relative momenta,
and the Coulomb interaction between them is taken into account 
at the level of experimental data, by factoring out the Gamow-Sommerfeld 
factors that stem from the $\pi^+d$ Coulomb interaction
--- see, e.g.~\cite{Heimberg}. 

\subsubsection{Pion production in the $s$-wave}
In order to calculate the amplitude of the $s$-wave pion production, that is,
the value of $C_0$, we took 
the result of Refs.~\cite{lensky2,Lensky:2006af}, where the amplitude of
the $s$-wave pion production 
was calculated up to NLO. We correct this amplitude by a constant factor so as
to get at threshold the 
value of $s$-wave production parameter $\alpha=252\mu\mathrm{b}$. This
quantity, $\alpha$, is the total cross section divided by the final pion
momentum in units of the pion mass, and the given value is extracted from the
width of pionic deuterium~\cite{Hauser,Strauch}.

\subsection{The reaction {\boldmath$pp\to pn\pi^+$}}
The transitions that contribute to the reaction $pp\to pn\pi^+$ at energies close to threshold
are ${^{1\!}}S_0\to{^3\!}S_1p,\ {^{1}\!}S_0\to{^3\!}D_1p$,
${^1\!}D_2\to{^3\!}S_1p,\ {^1\!}D_2\to{^3\!}D_1p$, and ${^3\!}P_1\to
{^3\!}S_1s$, ${^3\!}P_1\to {^3\!}D_1s$. 
At very low energies, the final ${^3\!}D_1$ state does not contribute,
however, the transitions 
to the ${^3\!}D_1$ state contribute via the ${^3\!}D_1\leftrightarrow{^3\!}S_1$
coupled channel. At these low energies, 
 the spin-angular structure of the amplitude reads
\begin{equation}
\maM_{pp\to pn\pi^+}=\bigg[\tilde{C}_0\,(\vec\mathcal{S}\times \hat
p\,)\,\vec\mathcal{S}\,'+\tilde C_1\mathcal{I}(\hat k\vec\mathcal{S}\,')+\tilde
C_2\,\mathcal{I} [(\hat p\hat k)(\hat p\vec\mathcal{S}\,')-\frac{1}{3}(\hat
k\vec\mathcal{S}\,')]\bigg], 
\end{equation}
where $\vec \mathcal{S}=\chi^T_2\frac{\sigma_2}{\sqrt{2}}\vec \sigma\chi_1$,
$\mathcal{I}=\chi^T_{2}\frac{\sigma_2}{\sqrt{2}}\chi_{1}$,  $\vec
\mathcal{S}\,'=\chi^\dagger_{1'}\vec
\sigma\frac{\sigma_2}{\sqrt{2}}\chi^*_{2'}$ are normalized spin structures
corresponding to the initial spin-triplet, initial spin-singlet, and final
spin-triplet states, in order. Here, $\tilde C_0$, $\tilde C_1$, and $\tilde
C_2$ are the amplitudes corresponding to the ${^3\!}P_1\to{^3\!}S_1s$,
${^1\!}S_0\to {^3\!}S_1p$, and 
${^{1}\!}D_2\to {^3\!}S_1p$ transitions, in order. Their relation with
the corresponding amplitudes in the $JLS$ basis is 
given by
\begin{eqnarray}
\tilde C_0&=&\sqrt{\frac{3}{2}}\ \maA^\mathrm{full}[{^3\!}P_1,{^3\!}S_1]\,,\\
\tilde C_1&=& \maA^\mathrm{full}[{^1\!}S_0,{^3\!}S_1] \,,\\
\tilde C_2&=&\sqrt{\frac{15}{2}}\ \maA^\mathrm{full}[{^1\!}D_2,{^3\!}S_1]\,.
\end{eqnarray}
Besides these amplitudes, there are also amplitudes that correspond to
the transition to the isospin-one ${^1\!}S_0$ final 
$pn$ state. However, these amplitudes do not interfere with $\tilde C_0$,
$\tilde C_1$, and $\tilde C_2$ because of 
different spins in the final state and generate only an additive part
in the cross section --- see also the discussion 
in the text. The observables in the reaction $pp\to pn\pi^+$ can be expressed
through $\tilde C_0$, $\tilde C_1$, and $\tilde C_2$ 
in the following way:
\begin{eqnarray}
\frac{d\sigma}{d\Omega} &=&\frac{1}{2(4\pi)^4M_N s
  p}\int\limits_0^{p_\mathrm{max}}k_\pi(p')p'^2dp'\bigg[2|\tilde
C_0|^2+|\tilde C_1|^2+\frac{1}{9}|\tilde C_2|^2(3\cos^2\theta_{\pi}+1)\nonumber\\ 
& &+\frac{2}{3}\mathrm{Re}\, (\tilde C_1\tilde
C_2^*)(3\cos^2\theta_{\pi}-1)\bigg]+\frac{d\sigma}{d\Omega}^{I=1}\\ 
A_y\cdot\frac{d\sigma}{d\Omega}&=&\frac{1}{2(4\pi)^4M_N s
  p}\int\limits_0^{p_\mathrm{max}}k_\pi(p')p'^2dp'\cdot 2 \sin\theta_{\pi}\cos\phi\
\mathrm{Im}\,\big(\tilde C_0^*(\tilde C_1-\frac{\tilde C_2}{3})\big). 
\end{eqnarray}
Here, $p_\mathrm{max}$ is the maximum relative momentum of the final nucleons, and
$\displaystyle\frac{d\sigma}{d\Omega}^{I=1}$ 
is the contribution of the final $pn$ state with isospin one to the
cross section. 

The expressions for the transition matrix elements relevant for the
calculation of $\tilde C_1$ and $\tilde C_2$ 
are the same as for the reaction $pp\to d\pi^+$, and therefore we 
refer the reader to the corresponding formulae, given in Sec.~\ref{reacdpi}
above. However, 
we should make a remark about the value of $\tilde C_0$. 
Instead of calculating it at NLO 
we  extracted  $\tilde C_0$ directly from the data ---
see the discussion in the main text.

\section{Resonance saturation vs. explicit Delta}
\label{saturation}

In the chiral limit the masses of both the $\Delta$ and the nucleon
stay finite and differ from each other. Thus, there is a well defined limit of QCD
where $m_\pi/\Delta M$, with $\Delta M=M_\Delta-M_N$, is a small
parameter. Consequently, the Delta degrees of freedom may be integrated
out and the effects of the $\Delta$ isobar are then absorbed into the 
LECs of the resulting Lagrangian. On the other hand, when the Delta 
degrees of freedom are included dynamically in the $NN$ system,
certain selection rules apply. In particular, the $\Delta$ is allowed
to contribute only, if the $NN$ system has the total isospin equal one.
It is instructive to discuss the implications of these selection rules 
for the reaction $NN\to NN\pi$ using both the EFT with and without explicit
Delta degrees of freedom. The corresponding discussion for $\pi N$
scattering can be found in Ref.~\cite{ulfbible}.

\begin{figure}[tb]
\begin{center}
\includegraphics[scale=0.5]{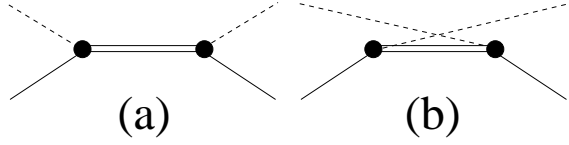}
\end{center}
\caption{\label{fig:piN}Leading diagrams
for $\pi N$ scattering with intermediate Delta.
Shown are the $s$-channel (a) and the $u$-channel (b)
contribution.}
\end{figure}

For simplicity, let us start from elastic $\pi N$
scattering. Using the interactions defined in the
main text one finds straightforwardly for diagram (a) of Fig.~\ref{fig:piN}
in the limit $\Delta M\to\infty$
\be \nonumber
iA_{(a)}^{ \pi \ {\rm elast}} &=& -i\left(\frac{2M_Nh_A^2}{4f_\pi^2\Delta M}\right)(\vec S \cdot \vec q\, ')
T_b (\vec S^\dagger \cdot \vec q) T_a^\dagger \\
&=& -i\left(\frac{M_Nh_A^2}{18f_\pi^2\Delta M}\right)
 q\, '_i q_j(2\delta_{ij}-i\epsilon_{ijk}\sigma_k)
(2\delta_{ba}-i\epsilon_{bac}\tau_c) \ ,
\label{aaelast}
\ee
where $a$ ($b$) and $\vec q$ ($\vec q\, '$) denote the isospin quantum number 
and momentum of the incoming (outgoing) pion.
Analogously, one finds for the $u$-channel diagram
\be \nonumber
iA_{(b)}^{ \pi \ {\rm elast}} &=& -i\left(\frac{2M_Nh_A^2}{4f_\pi^2\Delta M}\right)(\vec S \cdot \vec q)
T_a (\vec S^\dagger \cdot \vec q\, ') T_b^\dagger  \\
&=& -i\left(\frac{M_Nh_A^2}{18f_\pi^2\Delta M}\right)
 q\, '_i q_j(2\delta_{ij}-i\epsilon_{jik}\sigma_k)
(2\delta_{ba}-i\epsilon_{abc}\tau_c) \ .
\label{abelast}
\ee
In the second line of the above equation we used 
Eqs.~(\ref{deltanorm}). 
Thus, diagrams (a) and (b) are individually given by four terms. However,
two of them get canceled when adding the contributions together. The remaining
two terms in the sum exactly resemble the structure of the $c_3$
and $c_4$ terms of the effective Lagrangian.
One then finds for the Delta contribution
to the LECs $c_i$ the following result~\cite{ulfbible}:
\be
c_3^\Delta = -2c_4^\Delta = \frac{h_A^2}{9\Delta M} \ .
\ee
Obviously this matching is only possible after adding the two diagrams
together. On the other hand, the above mentioned selection rules for $NN\to NN\pi$
 are operative at the level of the individual diagrams. We now show how these
 two facts can be realized simultaneously.

In full analogy to the expressions given above, one finds
for the amplitudes corresponding to the diagrams of
Fig.~\ref{fig:NN2NNpi_direct} involving the Delta:
\be \nonumber
iA_{(c)} &=& -\left(\frac{M_N^2g_Ah_A^2}{18f_\pi^3\Delta M}\right)\tau^{(2)}_a
\left(\sigma^{(2)}\cdot \vec q\right)
\frac{1}{q^2-m_\pi^2}
 q\, '_i q_j\left(2\delta_{ij}-i\epsilon_{ijk}\sigma_k^{(1)}\right)
\left(2\delta_{ba}-i\epsilon_{bac}\tau_c^{(1)}\right) \\
\nonumber
&=& -\left(\frac{M_N^2g_Ah_A^2}{18f_\pi^3\Delta M}\right)\tau^{(2)}_a
\left(\sigma^{(2)}\cdot \vec q\right)
\frac{1}{q^2-m_\pi^2} q\, '_i q_j\\
& & \quad \times \left( \left[4\delta_{ij}\delta_{ba}-
\epsilon_{ijk}\sigma_k^{(1)}\epsilon_{bac}\tau_c^{(1)}\right]
-2i\left\{\epsilon_{ijk}\sigma_k^{(1)}\delta_{ab}+\delta_{ij}\epsilon_{bac}\tau_c^{(1)}
\right\}\right)\,,
\label{aaprod}
\\
\nonumber
iA_{(d)} &=& -\left(\frac{M_N^2g_Ah_A^2}{18f_\pi^3\Delta M}\right)\tau^{(2)}_a
\left(\sigma^{(2)}\cdot \vec q\right)
\frac{1}{q^2-m_\pi^2}
 q\, '_i q_j\left(2\delta_{ij}-i\epsilon_{jik}\sigma_k^{(1)}\right)
\left(2\delta_{ba}-i\epsilon_{abc}\tau_c^{(1)}\right) \\
\nonumber
&=& -\left(\frac{M_N^2g_Ah_A^2}{18f_\pi^3\Delta M}\right)\tau^{(2)}_a
\left(\sigma^{(2)}\cdot \vec q\right)
\frac{1}{q^2-m_\pi^2} q\, '_i q_j\\
& & \quad \times \left( \left[4\delta_{ij}\delta_{ba}-
\epsilon_{ijk}\sigma_k^{(1)}\epsilon_{bac}\tau_c^{(1)}\right]
+2i\left\{\epsilon_{ijk}\sigma_k^{(1)}\delta_{ab}+\delta_{ij}\epsilon_{bac}\tau_c^{(1)}
\right\}\right) \ ,
\label{abprod}
\ee
where in both amplitudes the momentum of the outgoing (virtual) pion is labelled as $q'$ ($q$).
In both amplitudes the terms in $[...]$  exhibit
the same spin-isospin structure as the $c_i$ terms
discussed above, while those in $\{...\}$
have a different structure.

Therefore, when diagram (c) and (d) are added together, only
the structures of the $c_i$ parameters survive.
However, due to the above mentioned selection 
rules, the diagrams contribute individually to different channels of $NN\to NN\pi$. 
To be specific, the diagram (d)  vanishes for the isospin-one to  
isospin-zero transition, i.~e.~for $pp\to (d/pn_{I=0})\pi^+$, 
whereas the diagram (c) does not. Furthermore, once the isospin matrix element
is evaluated for the diagram (c) it turns out 
that the expression in $\{...\}$ gives the same contribution as the one from
$[...]$.  
The same holds for  the isospin-zero to the isospin-one transition, i.~e.~for
$pn\to pp\pi^-$ ---  the diagram (c) does not contribute
whereas the terms in brackets $\{...\}$ and   $[...]$ for the diagram (d)
are equal. Therefore,
in the limit $\Delta M\to \infty$, one indeed observes
both properties simultaneously, namely that the
$N\Delta$ intermediate state does not contribute
if the external $NN$ state is in the isospin-zero state and
that the Delta effects can be absorbed in local
counter terms, namely $c_3$ and $c_4$.

We are now also in the position to see how the pattern changes 
when we start to move away from the limit $\Delta M\to \infty$.
Then the factors $1/\Delta M$ that appear in Eqs.
(\ref{aaelast})-(\ref{abprod}) should be replaced by the dynamical $N\Delta$
propagators.
One finds for the resulting combination of the propagators for both $\pi N$ scattering
and $NN\to NN\pi$
\be
\frac{1}{\Delta M-\omega_\pi}\pm \frac{1}{\Delta M+\omega_\pi} \ ,
\ee
where we used that $\omega_\pi \simeq E_{tot}$
and dropped terms significantly smaller than $m_\pi$.
In this expression the upper (lower) sign refers to
the combination of propagators relevant for the 
terms that can (cannot) be mapped onto the $c_i$.
Thus, the additional terms are suppressed by
\be
\delta = \omega_\pi/\Delta M \ .
\ee
Numerically $\delta$ is already as large as 0.5
at threshold and grows as one goes to higher
energies. Clearly, near the two-pion production threshold, $\delta\simeq 1$ and
it is necessary to keep the Delta as dynamical degrees of freedom.
See Ref.~\cite{vladimirdaniel} for a power counting that allows one to also
study the latter regime.


\begin{thebibliography}{0}
\bibitem{withkanzo}
 C.~Hanhart and K.~Nakayama,
  Phys.\ Lett.\  B {\bf 454} (1999) 176; for the discussion here especially
the appendix given only in arXiv:nucl-th/9809059 is relevant.
\end{thebibliography}
\end{document}